\documentclass[12pt]{article}
\usepackage[ascii]{inputenc}
\usepackage{amsmath,amssymb,amsfonts,amsthm}
\usepackage[margin=2.65cm]{geometry}
\usepackage{graphicx}
\usepackage{bbm}
\usepackage[caption=false]{subfig}
\usepackage[pdftex,bookmarks=false,colorlinks=true,linkcolor=blue,
citecolor=blue,filecolor=black,urlcolor=blue]{hyperref}

\providecommand{\llangle}{\langle\!\langle}
\providecommand{\rrangle}{\rangle\!\rangle}
\providecommand{\lllangle}{\langle\!\langle\!\langle}
\providecommand{\rrrangle}{\rangle\!\rangle\!\rangle}
\providecommand{\ud}{\mathrm{d}}
\numberwithin{equation}{section}

\begin{document}

\title{Low temperature dynamics of the one-dimensional discrete nonlinear Schr\"odinger equation}

\author{Christian B. Mendl
\footnote{Present address: Geballe Laboratory for Advanced Materials, Stanford University, 476 Lomita Mall, California 94305, USA}
$^\dagger$
and Herbert Spohn
\footnote{Zentrum Mathematik and Physik Department,
Technische Universit\"at M\"unchen,
Boltzmannstra{\ss}e 3, 85747 Garching bei M\"unchen, Germany.
Email: \href{mailto:mendl@ma.tum.de}{mendl@ma.tum.de}, \href{mailto:spohn@ma.tum.de}{spohn@ma.tum.de}}}

\date{May 12, 2015}

\maketitle

\begin{abstract}
We study equilibrium time correlations for the discrete nonlinear Schr\"odinger equation on a one-dimensional lattice and unravel three dynamical regimes. There is a high temperature regime with density and energy as the only two conserved fields. Their correlations have zero velocity and spread diffusively. In the low temperature regime umklapp processes are rare with the consequence that  phase differences appear as an additional (almost) conserved field. In an approximation where all umklapp is suppressed, while the equilibrium state remains untouched, one arrives at an anharmonic chain. Using the method of nonlinear fluctuating hydrodynamics we establish that the DNLS equilibrium time correlations have the same signature as a generic anharmonic chain, in particular KPZ broadening for the sound peaks and L\'evy 5/3 broadening for the heat peak. In the, so far not sharply defined, ultra-low temperature regime the integrability of the dynamics becomes visible. As an illustration we simulate the completely integrable Ablowitz-Ladik model and confirm ballistic broadening of the time correlations.
\end{abstract}

\newpage

\tableofcontents

\newpage

\section{Introduction}\label{sec1}

The one-dimensional discrete nonlinear Schr\"odinger equation (DNLS) is widely used as theoretical description of nonlinear optical wave guides \cite{CLS03}, semiclassical approximation to the dynamics of Bose-Einstein condensates \cite{HaFi14}, electronic transport in biomolecules \cite{Sc2003} and more, as discussed in the surveys \cite{Kevrekidis2009,FrLiOpPo11}. The DNLS has a surprisingly rich dynamical behavior. Most studies are concerned with finite energy solutions, which physically corresponds to zero temperature. In this article we focus on the dynamics at non-zero temperatures, a topic which lies in the domain of non-equilibrium statistical mechanics.

DNLS governs the dynamics of a complex-valued wave field denoted by $\psi_j$ with $j \in \mathbb{Z}$, the one-dimensional lattice, and is defined through the hamiltionian
\begin{equation}
\label{eq:nonintNLS_hamiltonian0}
H = \sum_{j\in \mathbb{Z}}\big( \tfrac{1}{2m} \lvert\psi_{j+1} - \psi_j\rvert^2 + \tfrac{1}{2}\,g\,\lvert\psi_j\rvert^4\big)\,.
\end{equation}
We will consider the defocusing case with $g > 0$. Kulkarni and Lamacraft \cite{KulkarniLamacraft2013} recently simulated the DNLS dynamics at density $\bar{\rho} = 1$, $m=1$, $g =1$, and temperature $\beta^{-1} = 0.005$. They discovered that the density-density time correlations in thermal equilibrium have symmetrically located sound peaks, which travel ballistically and broaden as $t^{2/3}$, consistent with the predictions based on the KPZ (alias stochastic Burgers) equation. As further amplified in \cite{Kulkarni2015}, one writes down a pair of coupled noisy Burgers equations for the left and right moving modes. Since the peaks separate linearly in time, it is argued that they decouple and each peak is governed by a scalar noisy Burgers equation, for which the exact scaling exponent and shape function is known \cite{PrSp2004, FeSp06, ImSa13, BCFV14}. In contrast, the high temperature DNLS has diffusive transport, as confirmed through molecular dynamics (MD) simulations of a steady state with stochastic boundary conditions forcing a flux of density and energy \cite{IubiniLepriPoliti2012}. Measured are the Onsager coefficients relating the linear response in the flux to the forcing. As we will argue, there is no sharp transition but the borderline between low and high temperatures is roughly characterized by
\begin{equation}\label{eq:lt_parameter_regime0}
\beta g \bar{\rho}^2 \sim 4\,,\qquad \tfrac{1}{m} \beta \bar{\rho} \sim 1 \,.
\end{equation}
Increasing the temperature even further one reaches the infinite temperature line, in dependence on the density, which borders the region of negative temperature states, realized through the appropriate microcanonical ensemble. There one observes interesting coarsening processes which are carried by the motion and coalescence of breathers \cite{IuPoPo13}.

In our contribution we will develop a global picture for positive temperature states, more tightly linked to DNLS than the previous discussions \cite{KulkarniLamacraft2013,Kulkarni2015}. The central dynamical concept are umklapp processes, at which the phase difference between neighboring sites crosses the values $\pm \pi$. At high temperatures umklapp happens frequently. Density and energy are the only conserved fields and their correlations spread diffusively with no systematic drift. However, in the equilibrium ensemble at low temperatures the phase differences are small, of the order $1/\sqrt{\beta}$. Umklapp is an activated dynamical process and occurs only with a frequency of the order $\mathrm{e}^{-\beta \Delta E}$, with $\Delta E$ a suitable energy barrier. As the temperature is lowered, the field of phase differences becomes almost conserved. The conservation law for phase difference degrades as $\mathrm{e}^{-t/\tau}$ with a life time $\tau$ proportional to $\mathrm{e}^{\beta \Delta E}$. For all practical purposes, in particular for MD simulations, the phase differences are locally conserved, once $\beta$ is in the domain with a border line specified by \eqref{eq:lt_parameter_regime0}. Thus one arrives at a system of three local conservation laws.

The described dynamical mechanism is familiar from second sound in liquid helium. There is no first sound for DNLS. But sound propagation is enabled through the appearance of an additional conserved field. In one spatial dimension such a transition is not sharp.

In fact there is a third regime, baptized ultra-low, whose precise border has still to be explored. In the low temperature regime there are three sharp peaks and for them we expect universal scaling laws with the dependence on the microscopic interactions lumped together in a few non-universal coefficients. In the ultra-low regime the integrability of the dynamics becomes visible. The density-density time correlations have now in addition, or instead, broader structures which self-similarly expand on a scale linear in time.

For the low temperature regime we elucidate the universal features, using the method of nonlinear fluctuating hydrodynamics. For this purpose our main tool is an effective low temperature hamiltonian, build such that the equilibrium ensemble remains unchanged while umklapp is completely suppressed. Hence phase difference, density, and energy are strictly conserved. In the well understood case of an anharmonic chain \cite{SpohnAHC2014}, there are also three conserved fields, stretch, momentum, and energy. The equilibrium state is of product form in the lattice index $j$. The DNLS hamiltonian contains however a nearest neighbor interaction, which complicates considerably the computation of the coupling constants for the normal mode representation used in nonlinear fluctuating hydrodynamics. Still we manage to ensure that the self-coupling of the heat mode vanishes, while the corresponding coefficients for the sound modes are different from zero. With this input the spreading and shape of the three peaks are predicted to be identical to the ones of a generic anharmonic chain, up to model-dependent non-universal coefficients.

In essence, our predictions are based on conservation laws and a sufficiently chaotic nonintegrable dynamics. With this view, one would expect that our results are also applicable to one-dimensional quantum systems. In fact, quite some time ago Andreev \cite{An80,Sa80} studied low temperature one-dimensional Fermi fluids and argued already that the sound peaks broaden as $t^{2/3}$, which turns out to be the same scaling exponent as for DNLS. We refer to \cite{PuZw05} for more recent studies. One-dimensional quantum fluids in the continuum are not readily accessible to quantum simulations. On the other hand, by quantizing the DNLS hamiltonian one arrives at the Bose-Hubbard model on a one-dimensional lattice. The propagating sound peaks are expected to be visible at low temperatures and density of order one, which are parameters more favorable for quantum simulations. Of course, more optimistically one would hope to be able to realize such propagating sound peaks in an cold atom experiment.

In view of the ultra-low temperature regime, we simulate the Ablowitz-Ladik model, which is an innocent modification of the DNLS dynamics, but known to be completely integrable. For this model, in a certain sense, every conserved mode generates a peak. Thus heuristically the correlations should be extended and widen ballistically.  For the continuum NLS the same time correlation structure is expected to appear. The underlying lattice is essential for our findings.

To provide a brief summary: First the high temperature regime is studied and simulated at $\beta = 1$, with all other parameters as before. We then transform to action-angle variables and determine the low temperature Hamiltonian. For it we compute the nonlinear coupling coefficients in the normal mode representation and determine the Landau-Placzek ratios. Thereby the predictions for the shape functions and the non-universal coefficients are made available, which are then compared with MD simulations of DNLS at $\beta = 15$. For the ultra-low temperature regime we set $\beta = 200$. In a final chapter we discuss the Ablowitz-Ladik model.

\section{One-dimensional DNLS, basic properties}\label{sec2}

Our starting point is the complex-valued field $\psi_j$ governed by the DNLS hamiltonian 
\begin{equation}
\label{eq:nonintNLS_hamiltonian}
H = \sum_{j=0}^{N-1}\big( \tfrac{1}{2m} \lvert\psi_{j+1} - \psi_j\rvert^2 + \tfrac{1}{2}\,g\,\lvert\psi_j\rvert^4\big)
\end{equation}
with mass $m > 0$ and coupling constant $g > 0$, i.e., a \emph{defocusing} nonlinearity. We impose periodic boundary conditions, $\psi_{N} = \psi_{0}$. Our interest will be in random initial data distributed according to a thermal Gibbs state, for which we will take the limit $N\to \infty$ at a suitable stage. The dynamics is defined through
\begin{equation}
\mathrm{i} \, \tfrac{\ud}{\ud t} \psi_j = \partial_{\psi_j^*} H\,,
\end{equation}
where $^*$ denotes complex conjugate. Then
\begin{equation}
\label{eq:nonintNLS}
\mathrm{i} \, \tfrac{\ud}{\ud t} \psi_j = -\tfrac{1}{2 m} \Delta \psi_{j} + g\,\lvert\psi_j\rvert^2\,\psi_j
\end{equation}
with the lattice Laplacian $\Delta = -\partial ^\mathrm{T}\partial$ and $\partial \psi_j = \psi_{j+1} - \psi_j$. The kinetic energy is chosen such that in the limit of zero lattice spacing one arrives at the continuum nonlinear Sch\"odinger equation. One could also expand the square, resulting in a hopping term plus a contribution proportional to the particle number,
\begin{equation}
\mathsf{N} = \sum_{j=0}^{N-1} \lvert\psi_j\rvert^2\,.
\end{equation}
 The sign of the hopping term does not play a role, since it can be flipped through the gauge transformation $\psi_j \mapsto \mathrm{e}^{\mathrm{i} \pi j} \psi_j $.

The DNLS has two obvious locally conserved fields, density and energy, 
\begin{equation}\label{eq:density_energy_def}
\rho_j = \lvert\psi_j\rvert^2\,,\qquad e_j = \tfrac{1}{2m} \lvert\psi_{j+1} - \psi_j\rvert^2 + \tfrac{1}{2}\,g\,\lvert\psi_j\rvert^4\,.
\end{equation}
According to the discussion in \cite{Ablowitz2004}, the DNLS is nonintegrable and one expects density and energy to be the only locally conserved fields. They satisfy the conservation laws
\begin{equation}
\begin{split}
\tfrac{\ud}{\ud t} \rho_j(t) + \mathcal{J}_{\rho,j+1}(t) - \mathcal{J}_{\rho,j}(t) &= 0\,,\\[1ex]
\tfrac{\ud}{\ud t} e_j(t) + \mathcal{J}_{e,j+1}(t) - \mathcal{J}_{e,j}(t) &= 0\,,
\end{split}
\end{equation}
with density current
\begin{equation}
\mathcal{J}_{\rho,j} = \tfrac{1}{2m} \mathrm{i} \big( \psi_{j-1}\,\partial \psi_{j-1}^* - \psi_{j-1}^*\,\partial\psi_{j-1} \big)
\end{equation}
and energy current
\begin{equation}
\mathcal{J}_{e,j} = \tfrac{1}{4 m^2} \mathrm{i} \big(\Delta\psi_{j}^* \,\partial\psi_{j-1} - \Delta\psi_{j}\,\partial\psi_{j-1}^* \big) + g \lvert\psi_j\rvert^2 \mathcal{J}_{\rho,j}\,.
\end{equation}

Canonically conjugate variables are introduced by splitting the wave function into its real and imaginary part as
\begin{equation}
\psi_j = \tfrac{1}{\sqrt{2}} (q_j + \mathrm{i} p_j)\,.
\end{equation}
In these variables, the hamiltonian reads
\begin{equation}
\label{eq:nonintNLS_hamiltonian_pq}
H = \sum_{j=0}^{N-1} \Big( \tfrac{1}{4 m} \big( (\partial q_j)^2 + (\partial p_j)^2 \big) + \tfrac{1}{8}\,g \big(q_j^2+p_j^2\big)^2 \Big)\,.
\end{equation}
The dynamics governed by Eq.~\eqref{eq:nonintNLS} is then identical to the hamiltonian system
\begin{equation}
\tfrac{\ud}{\ud t} q_j = \partial_{p_j}H\,, \quad \tfrac{\ud}{\ud t} p_j = -\partial_{q_j} H\,.
\end{equation}
Note that $H$ is symmetric under the interchange $q_j \leftrightarrow p_j$.

It will be convenient to make a canonical (symplectic) change of variables to polar coordinates as
\begin{equation}
\varphi_j = \mathrm{arctan}(p_j/q_j), \quad \rho_j = \tfrac{1}{2} \big(p_j^2+q_j^2\big)\,,
\end{equation}
which is equivalent to the representation
\begin{equation}
\psi_j = \sqrt{\rho_j}\, \mathrm{e}^{\mathrm{i} \varphi_j}\,.
\end{equation}
In the new variables the phase space becomes $(\rho_j, \varphi_j) \in \mathbb{R}_+ \times S^1$, with $S^1$ the unit circle. The corresponding hamiltonian is given by
\begin{align}
\label{eq:polarHamiltonian}
H &= \sum_{j=0}^{N-1} \Big( \tfrac{1}{2 m} \big( \sqrt{\rho_{j+1}\,\rho_j}\,2\,(1 - \cos(\varphi_{j+1} - \varphi_j)) + (\sqrt{\rho_{j+1}} - \sqrt{\rho_j})^2 \big) + \tfrac{1}{2}\,g\,\rho_j^2 \Big) \\
\label{eq:polarHamiltonian1}
&= \sum_{j=0}^{N-1} \big( {-}\tfrac{1}{m} \sqrt{\rho_{j+1}\,\rho_j}\, \cos(\varphi_{j+1} - \varphi_j) + \tfrac{1}{m} \rho_j + \tfrac{1}{2}\,g\,\rho_j^2 \big)\,.
\end{align}
The equations of motion read then
\begin{equation}\label{eq:motionDNLS}
\tfrac{\ud}{\ud t} \varphi_j = -\partial_{\rho_j} H\,, \quad \tfrac{\ud}{\ud t} \rho_j = \partial_{\varphi_j} H\,.
\end{equation}
From the continuity of $\psi_j(t)$ when moving through the origin, one concludes that at $\rho_j(t)= 0$ the phase jumps from $\varphi_j(t) $ to $ \varphi_j(t) + \pi$. The $\varphi_j$'s are angles and therefore position-like variables, while the $\rho_j$'s are actions and hence momentum-like variables. The hamiltonian depends only on phase differences which implies the invariance under the global shift $\varphi_j \mapsto \varphi_j + \phi$.

\section{Equilibrium time correlations at high temperatures}\label{sec3}

We consider the DNLS in thermal equilibrium, as described by the canonical ensemble
\begin{equation}\label{eq:ensembleDNLS}
Z_N(\mu,\beta)^{-1}\, \mathrm{e}^{-\beta(H - \mu \mathsf{N})} \prod_{j=-N/2}^{N/2-1} \ud\psi_j \ud\psi_j^*
\end{equation}
with inverse temperature $\beta$ and chemical potential $\mu$, $\beta > 0$, $\mu \in \mathbb{R}$. For the theory we take the limit $N \to \infty$. Numerically $N$ will be set to $4096$. Some of the derivations are done first at finite volume. Notationally we do not distinguish between finite $N$ and $N = \infty$, assuming that it will be understood from the context. Our interest is the propagation of small perturbations at large space-time scales, which are encoded by time correlations of the conserved fields, in our case density and energy,
\begin{equation}\label{eq:S_rho_e_def}
S_{\rho \rho}(j,t) = \langle \rho_j(t); \rho_0(0) \rangle\,, \quad S_{\rho e}(j,t) = \langle \rho_j(t);e_0(0)\rangle\,,\quad S_{ee}(j,t) = \langle e_j(t);e_0(0)
\rangle\,.
\end{equation}
Here $\langle \cdot \rangle$ denotes the average and $\langle \cdot;\cdot \rangle$ the second cumulant with respect to the canonical state \eqref{eq:ensembleDNLS}. $S(j,t)$ is defined on the entire lattice $\mathbb{Z}$. By invariance under reversal of time and space, $S_{\rho e}(j,t) = S_{e \rho}(j,t)$.

The canonical ensemble is even under complex conjugation of $\psi_j$, while the density and energy currents are of the form $\mathrm{i}(z - z^*)$. Hence
\begin{equation}
\langle \mathcal{J}_{\rho,j}\rangle = 0\,, \quad \langle \mathcal{J}_{e,j}\rangle = 0\,.
\end{equation}
If the dynamics is sufficiently chaotic, this property signals diffusive transport. More precisely we denote by $u_\rho$ the random deviation of the density from the uniform equilibrium density and by $u_e$ the one for the energy density. Then linear fluctuating hydrodynamics asserts that the random deviations are governed by the linear Langevin equation
\begin{equation}\label{eq:linearLangevin}
\partial_t u_\alpha +\partial_x\big(-\partial_x (D\vec{u}\,)_\alpha - (B\vec{\xi}\,)_\alpha\big)= 0\,,
\end{equation}
$\alpha = \rho,e$. Here $D$ is the $2\times 2$ diffusion matrix, $B$ the noise strength matrix, and $\xi_\alpha(x,t)$ are two independent normalized space-time white noises. We define the static covariance matrix through
\begin{equation}
\label{eq:sum_rule}
C_{\alpha\alpha'} = \sum_{j=-\infty}^{\infty} S_{\alpha\alpha'}(j,0) \,,
\end{equation}
from which it follows that $C = C^\mathrm{T}$, $C > 0$. The assumed fluctuation dissipation relation reads then
\begin{equation}
2 D C = B B^\mathrm{T}\,,
\end{equation}
which implies that the Onsager matrix $D C$ is a symmetric, strictly positive matrix. In particular $D$ has strictly positive eigenvalues. The covariance of the stationary process for \eqref{eq:linearLangevin} reads
\begin{equation}
\langle u_{\alpha}(x,t) u_{\alpha'}(0,0)\rangle = S_{\alpha\alpha'}(x,t) = \int_{\mathbb{R}} \ud k\, \mathrm{e}^{-\mathrm{i}2\pi k x} \big(\mathrm{e}^{-(2\pi k)^2 D\,t}\, C \big)_{\alpha\alpha'}\,.
\end{equation}
Working out the Fourier transform, one arrives at the prediction
\begin{equation}
\label{eq:S_heat_equation}
S(j, t) \simeq \tfrac{1}{\sqrt{4 \pi D\, t}} \, \mathrm{e}^{-j^2/(4 D\,t)} \,C\,.
\end{equation}
The square root of $D$ is unambiguous, since $D$ has strictly positive eigenvalues.

\begin{figure}[!ht]
\centering
\subfloat[density $S_{\rho \rho}(j,t)$]{
\includegraphics[width=0.3\textwidth]{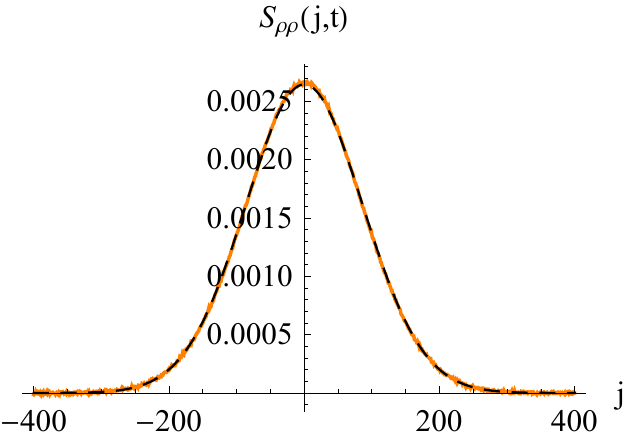}}
\hspace{0.02\textwidth}
\subfloat[density-energy $S_{\rho e}(j,t)$]{
\includegraphics[width=0.3\textwidth]{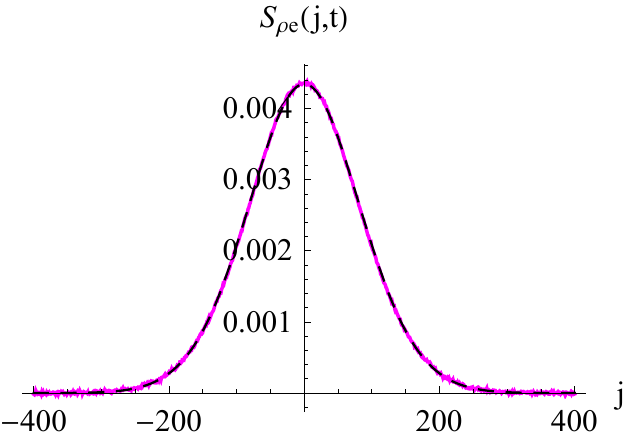}}
\hspace{0.02\textwidth}
\subfloat[energy $S_{e e}(j,t)$]{
\includegraphics[width=0.3\textwidth]{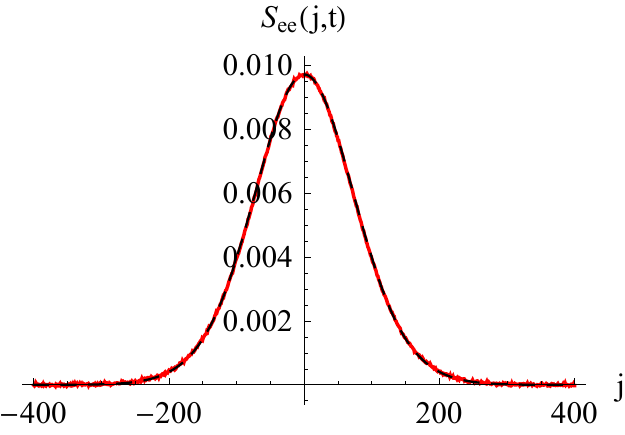}}
\caption{Equilibrium time-correlations $S_{\alpha \alpha'}(j,t)$ of the density and energy at high temperature, $\beta = 1$, for the discrete NLS with initial states drawn from the canonical ensemble \eqref{eq:ensembleDNLS}. The black dashed curves are the entries on the right of Eq.~\eqref{eq:S_heat_equation}.}
\label{fig:canonical_beta1}
\end{figure}

In Fig.~\ref{fig:canonical_beta1} we show the results of a MD simulation for $N=4096$, $m=1$, $g=1$, $\beta=1$, and $\langle \rho_j \rangle = 1$. The black dashed curves show the right side of \eqref{eq:S_heat_equation}, with $C$ determined by the sum rule \eqref{eq:sum_rule} and $D$ fitted numerically, under the constraint of $D C$ being symmetric. The numerical entries of $C$ are
\begin{equation}
C =
\begin{pmatrix}
0.580 & 0.907 \\
0.907 & 1.848 \\
\end{pmatrix} \,,
\end{equation}
and we record $D$ in the following table:
\begin{center}
\begin{tabular}{r|cccc}
$t$ & $256$ & $512$ & $1024$ & $1536$ \\
\hline
$D$ &
$\begin{pmatrix}
3.145 & -0.353 \\
2.509 &  0.822 \\
\end{pmatrix}$ &
$\begin{pmatrix}
3.137 & -0.364 \\
2.415 &  0.853 \\
\end{pmatrix}$ &
$\begin{pmatrix}
3.103 & -0.356 \\
2.350 &  0.876 \\
\end{pmatrix}$ &
$\begin{pmatrix}
3.079 & -0.350 \\
2.298 &  0.897 \\
\end{pmatrix}$
\end{tabular}
\end{center}
The values for $D$ drift only little in time, which supports the diffusive scaling exponent $\frac{1}{2}$, consistent with linear fluctuating hydrodynamics. In Fig.~\ref{fig:canonical_beta1} only the largest time, $t = 1536$, is shown.

\section{Dynamics at low temperatures}\label{sec4}

In the MD simulations \cite{KulkarniLamacraft2013} the parameters of the hamiltonian are the same as used here, but $\beta = 200$ at system size $N = 16384$. For the density-density correlations a behavior very different from the one reported in Sect.~\ref{sec3} is observed, in exhibiting ballistic propagation and non-diffusive spreading. A main goal of our contribution is to understand and predict such low temperature properties on the basis of nonlinear fluctuating hydrodynamics. The argument is involved and we choose to first illustrate our method for the case of coupled rotators (CR), for which the dependence on the analogue of $\rho_j$ is much simpler.

\paragraph{Coupled rotators.}
The CR hamiltonian reads
\begin{equation}
H_\mathrm{CR} = \sum_{j=0}^{N-1} \big(\tfrac{1}{2} p_j^2 - \cos(\varphi_{j+1} - \varphi_j)\big)\,.
\end{equation}
Here $\varphi_j \in S^1$ is the angle and $p_j \in \mathbb{R}$ the angular momentum of rotator $j$. We impose periodic boundary conditions as $\varphi_{N} = \varphi_0$. When compared to \eqref{eq:polarHamiltonian}, the prefactor $\tfrac{1}{m}\sqrt{\rho_{j+1}\,\rho_j}$ has been set equal to $1$ and the constraint $\rho_j \geq 0$ is ignored. Angular momentum and energy are the only locally conserved fields for $H_\mathrm{CR}$. Under the canonical equilibrium measure their average currents vanish as $N \to \infty$, signaling diffusive transport at high temperatures, as has been confirmed through MD simulations \cite{DasDhar2015, LLLHL2015}.

At zero temperature, there is the one-parameter family of ground states with $\varphi_j = \bar{\varphi}$, $p_j= 0$. When heating up, under the canonical equilibrium measure, the phase $\varphi_j$ jumps to $\varphi_{j+1}$ with a jump size $\mathcal{O}(1/\sqrt{\beta})$. For a more quantitative version, let us introduce the phase difference $\tilde{r}_j = \Theta(\varphi_{j+1} -\varphi_j)$, where $\Theta$ is $2\pi$-periodic and $\Theta(x) = x$ for $\lvert x \rvert \leq \pi$. Since $\Theta$ has a jump discontinuity, $\tilde{r}_j $ is not conserved. In a more pictorial language, the event that $\lvert\varphi_{j+1}(t) -\varphi_j(t)\rvert =\pi$ is called an umklapp for phase difference $\tilde{r_j}$ or an umklapp process to emphasize its dynamical character. At low temperatures a jump of size $\pi$ has a small probability of order $e^{-\beta\Delta V}$ with $\Delta V = 2$ the height of the potential barrier. Hence $\tilde{r}_j$ is locally conserved up to umklapp processes occurring with a very small frequency only, see \cite{DasDhar2015} for a numerical validation.

In the low temperature regime it is tempting to use an approximation, where the potential $V(x) = -\cos x$ is Taylor expanded at the minimum $x=0$. But such procedure would underestimate the regime of low temperatures, as can be seen from the example of a potential, still with $\Delta V = 2$, but several shallow minima. The proper small parameter is $\beta^{-1}$ such that $\beta \Delta V \gtrsim 2$, where $2$ is chosen to have a safety margin. To arrive at an optimal low temperature hamiltonian, we first parametrize the angles $\varphi_0, \dots, \varphi_{N-1}$ through $r_j = \varphi_{j+1} -\varphi_j$ with $r_j \in [-\pi,\pi]$. To distinguish, we denote the angles in this particular parametrization by $\phi_j$. The dynamics governed by $H_{\mathrm{CR}}$ corresponds to periodic boundary conditions at $r_j = \pm \pi$. For a low temperature description we impose instead specular reflection, i.e., if $r_j = \pm\pi$, then $p_j$, $p_{j+1}$ are scattered to $p_j' = p_{j+1}$, $p_{j+1}' = p_j$. By fiat all umklapp processes are now suppressed, while between two umklapp events the CR dynamics and the low temperature dynamics are identical. The corresponding hamiltonian reads
\begin{equation}
H_{\mathrm{CR,lt}} = \sum_{j=0}^{N-1} \big(\tfrac{1}{2} p_j^2 + \tilde{V}(\phi_{j+1} - \phi_j)\big)
\end{equation}
with
\begin{equation}
\tilde{V}(x) = -\cos x \ \ \text{for} \ \ \lvert x \rvert \leq \pi\,,\qquad \tilde{V}(x) = \infty \ \ \text{for}\ \ \lvert x \rvert > \pi\,.
\end{equation}
As before, periodic boundary conditions $\phi_{N} = \phi_0$ are understood. The pair $(\phi_j,p_j)$ are canonically conjugate variables. Note that as weights $\exp[-\beta H_{\mathrm{CR}}] = \exp[-\beta H_{\mathrm{CR,lt}}]$. Thus all equilibrium properties of the coupled rotators remain untouched.

The hamiltonian $H_\mathrm{CR,lt}$ is a variant of the hard collision model with square well potential, see \cite{MeSp14}, thus a conventional anharmonic chain. The dynamics governed by $H_\mathrm{CR,lt}$ has three conserved fields, the stretch $r_j = \phi_{j+1} - \phi_j$, the momentum $p_j$ and the energy $e_j = \tfrac{1}{2}p_j^2 + \tilde{V}(r_j)$. To study its equilibrium time correlations the methods and results from \cite{SpohnAHC2014} apply directly. Because of $\phi_0 = \phi_{N}$, one has $\sum_{j=0}^{N-1} r_j = 0$. The model is in the dynamical phase characterized by an even potential at zero pressure.

We claim that, for $\beta \Delta V \gtrsim 2$, the CR equilibrium time correlations are well approximated by those of $H_\mathrm{CR,lt}$, provided the time of comparison is not too long. The latter correlations can be obtained within the framework of nonlinear fluctuating hydrodynamics. Thereby one arrives at fairly explicit dynamical predictions for the low temperature regime of the CR model.

\paragraph{Low temperature approximation for DNLS.}
We return to the DNLS hamiltonian and restrict ourselves to the case $\mu > 0$. As can be inferred from the hamiltonian \eqref{eq:polarHamiltonian}, in the limit $\beta \to \infty$ the canonical measure converges to the one-parameter family of ground states with $\rho_j = \bar{\rho} = \mu/g$, $\varphi_j = \bar{\varphi}$ and $\bar{\varphi}$ uniformly distributed on $S^1$. As for CR, we introduce $\tilde{r}_j = \Theta(\varphi_{j+1} -\varphi_j)$. In the context of Bose-Einstein condensates $\tilde{v}_j = \tfrac{1}{m} \tilde{r}_j$ is called the superfluid velocity. At low temperatures, $\rho_j-\bar{\rho}$ is confined by an essentially harmonic potential, hence $\lvert\rho_j-\bar{\rho}\rvert = \mathcal{O}(1/\sqrt{\beta}\,)$ and the phase $\varphi_j$ jumps to $\varphi_{j+1}$ with a jump size $\mathcal{O}(1/\sqrt{\beta}\,)$, also.

The low temperature hamiltonian is constructed in such a way that the equilibrium ensemble remains unchanged while all umklapp processes are suppressed. To achieve our goal we follow verbatim the CR blueprint. Now $(\phi_j,\rho_j)$ are a pair of canonically conjugate variables, only $\rho_j \geq 0$ instead of $p_j \in \mathbb{R}$. The phases are parametrized such that $\varphi_{j+1} - \varphi_j$ lies in the interval $[-\pi,\pi]$. Umklapp is a point at the boundary of this interval. Thus the proper low temperature hamiltonian reads
\begin{equation}
H_{\mathrm{lt}} = \sum_{j=0}^{N-1} \big( \sqrt{\rho_{j+1}\,\rho_j} \, U(\phi_{j+1} - \phi_j) + V(\rho_j) \big)\,,
\end{equation}
where
\begin{equation}
U(x) = -\tfrac{1}{m} \cos(x) \ \ \text{for}\ \ \lvert x \rvert \leq \pi \,,\qquad U(x) = \infty \ \ \text{for}\ \ \lvert x \rvert > \pi\,,
\end{equation}
and 
\begin{equation}
V(x) = \tfrac{1}{m} x + \tfrac{1}{2} g\, x^2 \ \ \text{for}\ \ x \geq 0\,,\qquad V(x) = \infty \ \ \text{for}\ \ x < 0\,.
\end{equation}
For some computations it will be convenient to replace the hard collision potentials $U$, $V$ by a smooth variant, for which the infinite step is replaced by a rapidly diverging smooth potential. The dynamics is governed by
\begin{equation}\label{eq:motionHlt}
\tfrac{\ud}{\ud t} \phi_j = - \partial_{\rho_j} H_{\mathrm{lt}}\,,\qquad\tfrac{\ud}{\ud t} \rho_j = \partial_{\phi_j} H_{\mathrm{lt}}\,,
\end{equation}
including the specular reflection of $\rho_j$ at $\rho_j= 0$ and of $r_j$ at $r_j = \pm \pi$. As for CR, between two umklapp events the true and the low temperature dynamics are identical. Also as weights $\exp[-\beta H] = \exp[-\beta H_{\mathrm{lt}}]$, which is a salient feature of our approximation.

There are two potential barriers, $\Delta U$ and $\Delta V$. The minimum of $V(x) -\tfrac{1}{m}x - \mu x$ is at $\bar{\rho} = \mu/g$, hence $\Delta V = \tfrac{1}{2}g\bar{\rho}^2$. The minimum of $U$ is at $\phi_{j+1} - \phi_j = 0$ and, setting $\rho_j = \bar{\rho}$, one arrives at $\Delta U = \tfrac{2}{m} \bar{\rho}$. Thus the low temperature regime is characterized by
\begin{equation}\label{eq:lt_parameter_regime}
\tfrac{1}{2} \beta \bar{\rho}^2 \gtrsim 2\,,\qquad \tfrac{2}{m} \beta \bar{\rho} \gtrsim 2\,.
\end{equation}
In this parameter regime we expect the equilibrium time correlations based on $H_{\mathrm{lt}}$ to well approximate the time correlations of the exact DNLS.

As for a conventional anharmonic chain we introduce the stretch $r_j = \phi_{j+1} - \phi_j$. Then the conserved fields are $\rho_j$, $r_j$, and the energy
\begin{equation}
e_j = \sqrt{\rho_{j+1}\,\rho_j}\,U(r_j) + V(\rho_j) \,.
\end{equation}
This local energy differs from the one introduced in \eqref{eq:density_energy_def} by the term $\tfrac{1}{2m}(\rho_{j+1} - \rho_j)$. In the expressions below such a difference term drops out and in the final result we could use either one. The local conservation laws and their currents read, for the density
\begin{equation}
\tfrac{\ud}{\ud t} \rho_j + \mathcal{J}_{\rho,j+1} - \mathcal{J}_{\rho,j} = 0
\end{equation}
with local density current
\begin{equation}
\label{eq:current_z}
\mathcal{J}_{\rho,j} = \sqrt{\rho_{j-1}\,\rho_{j}} \, U'(r_{j-1})\,,
\end{equation}
for the stretch
\begin{equation}
\tfrac{\ud}{\ud t} r_j + \mathcal{J}_{r,j+1} - \mathcal{J}_{r,j} = 0
\end{equation}
with local stretch current
\begin{equation}
\label{eq:current_r}
\mathcal{J}_{r,j} = \tfrac{1}{2} \sqrt{\rho_{j+1}/\rho_j}\,U(r_j) + \tfrac{1}{2} \sqrt{\rho_{j-1}/\rho_j}\,U(r_{j-1}) + V'(\rho_j)\,,
\end{equation}
and for the energy
\begin{equation}
\tfrac{\ud}{\ud t} e_j + \mathcal{J}_{e,j+1} - \mathcal{J}_{e,j} = 0
\end{equation}
with local energy current
\begin{equation}
\label{eq:current_energy}
\mathcal{J}_{e,j} = \tfrac{1}{2} \sqrt{\rho_{j-1}\,\rho_{j+1}}\,\big( U(r_{j-1})U'(r_j) + U'(r_{j-1}) U(r_j) \big) + \sqrt{\rho_{j-1}\,\rho_j}\,U'(r_{j-1}) V'(\rho_j) \, .
\end{equation}
To shorten notation, we set $\vec{g}_j = (\rho_j,r_j,e_j)$ and $\vec{\mathcal{J}}_j = \big( \mathcal{J}_{\rho,j}, \mathcal{J}_{r,j}, \mathcal{J}_{e,j} \big)$.

To stress again, the low temperature hamiltonian $H_\mathrm{lt}$ serves only as an input to nonlinear fluctuating dynamics. Our MD simulations will use DNLS throughout.

\section{The coupling coefficients for fluctuating hydrodynamics}\label{sec5}

To arrive at predictions for the dynamics based on $H_\mathrm{lt}$ we will adopt the method from \cite{SpohnAHC2014}. In contrast to a standard anharmonic chain, the hamiltonian, written in terms of stretches, is no longer a sum of one-particle terms. At first sight this seems to complicate matters considerably. But, there is a surprising identity for the three average currents, which allows us to still obtain the nonlinear coupling coefficients in a concise form.

\paragraph{The case of general $\mu,\nu$.}
The chemical potential $\mu$ is the dual variable for $\rho_j$ and we introduce $\nu$ as the dual variable for $r_j$. Since $\phi_0 = \phi_N$, one has $\sum_{j=0}^{N-1} r_j = 0$. Hence, by the equivalence of ensembles, the equilibrium state is actually at $\nu = 0$. But for the quadratic expansion of the Euler currents it is more convenient to first work with general $\nu$, setting $\nu=0$ at the end. The canonical ensemble of $H_{\mathrm{lt}}$ for a finite system with $N$ lattice sites and periodic boundary conditions is given by
\begin{equation}
\label{eq:ensembleDNLS_lt}
Z_N(\mu,\nu,\beta)^{-1}\, \mathrm{e}^{-\beta \left( H - \mu \sum_{j=0}^{N-1} \rho_j - \nu \sum_{j=0}^{N-1} r_j \right)} \prod_{j=0}^{N-1} \ud \rho_j\, \ud r_j\, 
\end{equation}
with the normalizing partition function
\begin{equation}
Z_N(\mu,\nu,\beta) = \int_{({\mathbb{R_+}\times[-\pi,\pi]})^N} \mathrm{e}^{-\beta \left( H - \mu \sum_{j=0}^{N-1} \rho_j - \nu \sum_{j=0}^{N-1} r_j \right)} \prod_{j=0}^{N-1} \ud \rho_j\, \ud r_j\,.
\end{equation}
On purpose, the index ``lt'' has been omitted, since $H$ and $H_{\mathrm{lt}}$ define the same measure. The canonical free energy is defined as
\begin{equation}
\label{eq:F_def}
F(\mu,\nu,\beta) = - \beta^{-1} \lim_{N \to\infty} \tfrac{1}{N} \log Z_N(\mu,\nu,\beta)\,.
\end{equation}
The averages of $\rho_j$, $r_j$, $e_j$ are
\begin{equation}
\label{eq:avr_derivF}
\begin{split}
\rho &= \langle \rho_j \rangle = - \partial_{\mu} F(\mu,\nu,\beta), \qquad \mathsf{r} = \langle r_j \rangle = - \partial_{\nu} F(\mu,\nu,\beta), \\[1ex]
\mathsf{e} &= \lim_{N \to \infty} \tfrac{1}{N} \langle H \rangle = \partial_{\beta} (\beta\,F(\mu,\nu,\beta)) + \mu\,\rho + \nu\,\mathsf{r}\,,
\end{split}
\end{equation}
independent of $j$ by translation invariance. By convexity of $F$, these relations define the inverse mapping $(\rho,\mathsf{r},\mathsf{e}) \mapsto (\mu(\rho,\mathsf{r},\mathsf{e}), \nu(\rho,\mathsf{r},\mathsf{e}), \beta(\rho,\mathsf{r},\mathsf{e}))$. As discussed in Appendix~\ref{sec:average_currents}, the average currents turn out to be
\begin{equation}
\label{eq:current_avr}
\big\langle \vec{\mathcal{J}_j} \big\rangle = \langle ( \mathcal{J}_{\rho,j}, \mathcal{J}_{r,j}, \mathcal{J}_{e,j} ) \rangle = (\nu, \mu, \mu\,\nu) = \vec{\mathsf{j}}\,.
\end{equation}
The product form of the energy current will be crucial in what follows.

Taking derivatives with respect to $\mu$, $\nu$, $\beta$ generates sums over $j$. We therefore introduce the shorthand
\begin{equation}
\llangle f_0; h_0 \rrangle = \sum_{j=0}^{N-1} \langle f_j; h_0 \rangle
\end{equation}
with $\langle \cdot; \cdot \rangle$ denoting the second cumulant. Here $f_0$ refers to some local function and $f_j$ is the same function shifted by $j$. At the end of the computation one wants to take $N \to \infty$. Infinite volume correlation functions, such as $\langle f_j; h_0 \rangle = \langle f_j\,h_0 \rangle - \langle f_j \rangle \langle h_0 \rangle$, decay exponentially fast to zero. Hence all our formulas hold also for infinite volume. Note that in this case
\begin{equation}
\llangle f_0; h_0 \rrangle = \sum_{j=-\infty}^{\infty} \langle f_j; h_0 \rangle\,.
\end{equation}
With such notation the static susceptibility matrix is defined through
\begin{equation}
C_{\alpha\alpha'} = \llangle g_{\alpha,0}; g_{\alpha',0} \rrangle\,,\quad 
C = \begin{pmatrix}
\llangle \rho_0; \rho_0 \rrangle & \llangle \rho_0; r_0 \rrangle & \llangle \rho_0; e_0 \rrangle \\
\llangle r_0; \rho_0 \rrangle & \llangle r_0; r_0 \rrangle & \llangle r_0; e_0 \rrangle \\
\llangle e_0; \rho_0 \rrangle & \llangle e_0; r_0 \rrangle & \llangle e_0; e_0 \rrangle
\end{pmatrix} \,,
\end{equation}
$\alpha, \alpha' = 1, 2, 3$, where the vector $\vec{g}_j$ of the field variables is defined below Eq.~\eqref{eq:current_energy}.

The following relations hold,
\begin{equation}
\label{eq:davr_cumulant2}
\partial_{\mu} \langle f_0 \rangle = \beta \llangle f_0; \rho_0 \rrangle\,, \quad \partial_{\nu} \langle f_0 \rangle = \beta \llangle f_0; r_0 \rrangle\,, \quad
\partial_{\beta} \langle f_0 \rangle = - \llangle f_0; e_0 - \mu\, \rho_0- \nu\, r_0 \rrangle \,.
\end{equation}
In matrix notation,
\begin{equation}
\label{eq:davr_cumulant2_matrix}
\begin{pmatrix}
\partial_{\mu}\\ \partial_{\nu}\\ \partial_{\beta}
\end{pmatrix} \langle f_0 \rangle= 
\begin{pmatrix}\beta & 0 & 0\\ 0 & \beta & 0 \\ \mu & \nu & -1 \end{pmatrix} \begin{pmatrix} \llangle f_0; \rho_0 \rrangle \\ \llangle f_0; r_0\rrangle \\ \llangle f_0; e_0 \rrangle \end{pmatrix} \,.
\end{equation}
Hence, using \eqref{eq:current_avr}, one deduces that
\begin{equation}
\label{eq:current_field_corr}
\llangle \mathcal{J}_{\alpha,0}; g_{\alpha',0} \rrangle = \frac{1}{\beta} \begin{pmatrix} 0 & 1 & \nu \\ 1 & 0 & \mu \\ \nu & \mu & 2 \mu\,\nu \end{pmatrix}\,.
\end{equation}

Setting
\begin{equation}
\llangle f_0; h_0; \ell_0 \rrangle = \sum_{j=0}^{N-1} \langle f_0; h_0; \ell_j \rangle\,,\qquad
\lllangle f_0; h_0; \ell_0 \rrrangle = \sum_{i,j=0}^{N-1} \langle f_0; h_i; \ell_j \rangle
\end{equation}
for the third cumulant, one obtains
\begin{equation}
\label{eq:dcumulant2_cumulant3}
\begin{split}
\partial_{\mu} \langle f_0; h_0 \rangle &= \beta \llangle f_0; h_0; \rho_0 \rrangle\,, \quad 
\partial_{\nu} \langle f_0; h_0 \rangle = \beta \llangle f_0; h_0; r_0 \rrangle\,, \\
\partial_{\beta} \langle f_0; h_0 \rangle &= - \llangle f_0; h_0; e_0 - \mu\, \rho_0 - \nu\, r_0 \rrangle \,.
\end{split}
\end{equation}
Employing \eqref{eq:current_field_corr}, one arrives at
\begin{align}
\label{eq:J1_cumulant3}
\mathcal{H}^{\rho}_{\alpha \gamma} = \lllangle \mathcal{J}_{1,0}; g_{\alpha,0}; g_{\gamma,0} \rrrangle &= \frac{1}{\beta^2} \begin{pmatrix} 0 & 0 & 0 \\ 0 & 0 & 1 \\ 0 & 1 & 2\nu \end{pmatrix}, \\
\label{eq:J2_cumulant3}
\mathcal{H}^{\mathsf{r}}_{\alpha \gamma} = \lllangle \mathcal{J}_{2,0}; g_{\alpha,0}; g_{\gamma,0}\rrrangle
 &= \frac{1}{\beta^2} \begin{pmatrix} 0 & 0 & 1 \\ 0 & 0 & 0 \\ 1 & 0 & 2\mu \end{pmatrix}, \\
\label{eq:J3_cumulant3}
\mathcal{H}^{\mathsf{e}}_{\alpha \gamma} = \lllangle \mathcal{J}_{3,0}; g_{\alpha,0}; g_{\gamma,0}\rrrangle &= \frac{1}{\beta^2} \begin{pmatrix} 0 & 1 & 2 \nu \\ 1 & 0 & 2 \mu \\ 2 \nu & 2 \mu & 6 \mu\,\nu \end{pmatrix}\,.
\end{align}

To write down the equations of nonlinear fluctuating hydrodynamics, one needs the average currents \eqref{eq:current_avr} expanded to second order at the uniform background values $(\bar{\rho}, \bar{\mathsf{r}}, \bar{\mathsf{e}})$. To streamline our formulas, we denote the deviation from the background as $\vec{u} = (u_1,u_2,u_3)$ $= (\rho - \bar{\rho}, \mathsf{r} - \bar{\mathsf{r}}, \mathsf{e} - \bar{\mathsf{e}})$. The background values will be suppressed in our notation. To linear order,
\begin{equation}
A_{\alpha \alpha'} = \partial_{u_{\alpha'}} \mathsf{j}_\alpha = \begin{pmatrix}
\partial_{\rho} \nu & \partial_{\mathsf{r}} \nu & \partial_{\mathsf{e}} \nu \\
\partial_{\rho} \mu & \partial_{\mathsf{r}} \mu & \partial_{\mathsf{e}} \mu \\
\partial_{\rho} (\mu\,\nu) & \partial_{\mathsf{r}} (\mu\,\nu) & \partial_{\mathsf{e}} (\mu\,\nu)
\end{pmatrix}.
\end{equation}
We use the identity
\begin{equation}
\label{eq:derivative_product}
\begin{pmatrix}
\partial_{\mu} \rho & \partial_{\mu} \mathsf{r} & \partial_{\mu} \mathsf{e} \\
\partial_{\nu} \rho & \partial_{\nu} \mathsf{r} & \partial_{\nu} \mathsf{e} \\
\partial_{\beta} \rho & \partial_{\beta} \mathsf{r} & \partial_{\beta} \mathsf{e} \\
\end{pmatrix}
\begin{pmatrix}
\partial_{\rho} \mu & \partial_{\rho} \nu & \partial_{\rho} \beta \\
\partial_{\mathsf{r}} \mu & \partial_{\mathsf{r}} \nu & \partial_{\mathsf{r}} \beta \\
\partial_{\mathsf{e}} \mu & \partial_{\mathsf{e}} \nu & \partial_{\mathsf{e}} \beta \\
\end{pmatrix} = \mathbbm{1}
\end{equation}
to express derivatives of $\mu$, $\nu$, $\beta$ by derivatives of $\rho$, $\mathsf{r}$, $\mathsf{e}$.
Applying to the left matrix the relation \eqref{eq:davr_cumulant2_matrix} leads to
\begin{equation}
\begin{pmatrix}
\partial_{\mu} \rho & \partial_{\mu} \mathsf{r} & \partial_{\mu} \mathsf{e} \\
\partial_{\nu} \rho & \partial_{\nu} \mathsf{r} & \partial_{\nu} \mathsf{e} \\
\partial_{\beta} \rho & \partial_{\beta} \mathsf{r} & \partial_{\beta} \mathsf{e}
\end{pmatrix} = \begin{pmatrix}\beta & 0 & 0\\ 0 & \beta & 0 \\ \mu & \nu & -1 \end{pmatrix} C.
\end{equation}
By matrix inversion, one thus obtains the right matrix in Eq.~\eqref{eq:derivative_product} as
\begin{equation}
\label{eq:coordJacobi_formula}
\begin{pmatrix}
\partial_{\rho} \mu & \partial_{\rho} \nu & \partial_{\rho} \beta \\
\partial_{\mathsf{r}} \mu & \partial_{\mathsf{r}} \nu & \partial_{\mathsf{r}} \beta \\
\partial_{\mathsf{e}} \mu & \partial_{\mathsf{e}} \nu & \partial_{\mathsf{e}} \beta \\
\end{pmatrix} =
\frac{1}{\beta}\, C^{-1} \begin{pmatrix}
1 & 0 & 0 \\
0 & 1 & 0 \\
\mu & \nu & -\beta \\
\end{pmatrix} \,.
\end{equation}
Furthermore
\begin{equation}
\label{eq:Aformula}
A = \frac{1}{\beta} \begin{pmatrix} 0 & 1 & \nu \\ 1 & 0 & \mu \\ \nu & \mu & 2 \mu\,\nu \end{pmatrix} C^{-1},
\end{equation}
where we have used that $C$ is symmetric. In particular, we confirm the general relations
\begin{equation}
A\,C = C\,A^{\mathrm{T}}\,,\qquad \llangle \mathcal{J}_{\alpha,0}; g_{\alpha',0} \rrangle = (A C)_{\alpha \alpha'} \,,
\end{equation}
compare with Appendix~1f of \cite{SpohnAHC2014}.

For the second order expansion we start from the chain rule, 
\begin{equation}
\begin{pmatrix} \partial_{\rho} f\\ \partial_{\mathsf{r}} f\\ \partial_{\mathsf{e}} f \end{pmatrix}
= \begin{pmatrix}
\partial_{\rho} \mu & \partial_{\rho} \nu & \partial_{\rho} \beta \\
\partial_{\mathsf{r}} \mu & \partial_{\mathsf{r}} \nu & \partial_{\mathsf{r}} \beta \\
\partial_{\mathsf{e}} \mu & \partial_{\mathsf{e}} \nu & \partial_{\mathsf{e}} \beta \\
\end{pmatrix}
\begin{pmatrix} \partial_{\mu} f\\ \partial_{\nu} f\\ \partial_{\beta} f \end{pmatrix} .
\end{equation}
The Jacobi matrix on the right can be obtained from Eq.~\eqref{eq:coordJacobi_formula}, and together with Eqs.~\eqref{eq:davr_cumulant2_matrix} and \eqref{eq:dcumulant2_cumulant3}, one arrives at
\begin{equation}
\begin{pmatrix} \partial_{\rho} \langle f_0 \rangle \\ \partial_{\mathsf{r}} \langle f_0 \rangle \\ \partial_{\mathsf{e}} \langle f_0 \rangle \end{pmatrix} = C^{-1} \begin{pmatrix} \llangle f_0; \rho_0 \rrangle \\ \llangle f_0; r_0 \rrangle \\ \llangle f_0; e_0\rrangle \end{pmatrix}
\quad\text{and}\quad
\begin{pmatrix} \partial_{\rho} \langle f_0; h_0 \rangle \\ \partial_{\mathsf{r}} \langle f_0; h_0 \rangle \\ \partial_{\mathsf{e}} \langle f_0; h_0 \rangle \end{pmatrix} = C^{-1} \begin{pmatrix} \llangle f_0; h_0; \rho_0 \rrangle \\ \llangle f_0; h_0; r_0 \rrangle \\ \llangle f_0; h_0; e_0 \rrangle \end{pmatrix} \,.
\end{equation}
In index notation,
\begin{equation}
\partial_{u_{\alpha}} \langle f_0 \rangle = \sum_{\alpha'=1}^3 (C^{-1})_{\alpha \alpha'} \, \llangle f_0; g_{\alpha',0} \rrangle
\quad\text{and}\quad
\partial_{u_{\alpha}} \langle f_0; h_0 \rangle = \sum_{\alpha'=1}^3 (C^{-1})_{\alpha \alpha'} \, \llangle f_0; h_0; g_{\alpha',0} \rrangle\,.
\end{equation}
This relation allows us to obtain second derivatives as
\begin{equation}
\partial_{u_{\alpha}} \partial_{u_{\gamma}} \langle f_0 \rangle =\sum_{\alpha',\gamma'=1}^{3} (C^{-1})_{\alpha \alpha'} (C^{-1})_{\gamma \gamma'} \, \lllangle f_0; g_{\alpha',0}; g_{\gamma',0} \rrrangle \,.
\end{equation}
In particular, according to Eqs.~\eqref{eq:J1_cumulant3} -- \eqref{eq:J3_cumulant3}, we obtain the Hessians of the currents
\begin{equation}
\label{eq:current_Hessian}
H^{\alpha}_{\gamma \gamma'} = \partial_{u_{\gamma}} \partial_{u_{\gamma'}} \mathsf{j}_\alpha = \big(C^{-1}\, \mathcal{H}^{\alpha}\, C^{-1}\big)_{\gamma \gamma'} \,.
\end{equation}
As required, the matrices on the right are symmetric, since $C$ is symmetric.

A matrix $R$ is introduced in \cite{SpohnAHC2014} for the transformation to normal modes. Specifically, $R$ diagonalizes $A$ and satisfies
\begin{equation}
\label{eq:Rproperties}
R A R^{-1} = \mathrm{diagonal}, \quad R C R^{\mathrm{T}} = \mathbbm{1}\,.
\end{equation}
Inserting Eq.~\eqref{eq:current_Hessian} into the definition of the coupling matrices,
\begin{equation}
G^{\alpha} = \tfrac{1}{2} \sum_{\alpha'=1}^3 R_{\alpha \alpha'}\, R^{-\mathrm{T}} H^{\alpha'} R^{-1} \,,
\end{equation}
and using that $C^{-1} R^{-1} = R^{\mathrm{T}}$ leads to
\begin{equation}
\label{eq:Gformula}
G^{\alpha} = \tfrac{1}{2} \sum_{\alpha'=1}^3 R_{\alpha \alpha'}\, R\, \mathcal{H}^{\alpha'} R^{\mathrm{T}} \,.
\end{equation}

\paragraph{Special case $\nu = 0$.}
The next step is to take $\nu = 0$, which implies $\mathsf{r} = \langle r_j \rangle = 0$, $\langle r_j; \rho_0 \rangle = 0$, and $\langle r_j; e_0 \rangle = 0$. The inverse matrix of $C$ simplifies to
\begin{equation}
C^{-1} = \frac{1}{\Gamma} \begin{pmatrix}
\llangle e_0; e_0 \rrangle & 0 & -\llangle \rho_0; e_0 \rrangle \\
0 & 0 & 0 \\
-\llangle \rho_0; e_0 \rrangle & 0 & \llangle \rho_0; \rho_0 \rrangle
\end{pmatrix} +
\frac{1}{\llangle r_0; r_0 \rrangle} \begin{pmatrix}
0 & 0 & 0 \\
0 & 1 & 0 \\
0 & 0 & 0
\end{pmatrix}
\end{equation}
with
\begin{equation}
\Gamma = \llangle \rho_0; \rho_0 \rrangle \llangle e_0; e_0 \rrangle - \llangle \rho_0; e_0 \rrangle^2 \,.
\end{equation}
Note that $\Gamma$ is invariant under $e_0 \to e_0 - \mu\,\rho_0$.

Using Eq.~\eqref{eq:Aformula} for $A$ at $\nu = 0$ results in
\begin{equation}
A = \frac{1}{\beta\, \Gamma} \begin{pmatrix}
0 & 0 & 0 \\
\llangle e_0 - \mu\,\rho_0; e_0 \rrangle & 0 & -\llangle \rho_0; e_0 - \mu\,\rho_0 \rrangle \\
0 & 0 & 0
\end{pmatrix} +
\frac{1}{\beta\, \llangle r_0; r_0 \rrangle} \begin{pmatrix}
0 &  1  & 0 \\
0 &  0  & 0 \\
0 & \mu & 0
\end{pmatrix} \,.
\end{equation}
The eigenvalues of $A$ are $(-c, 0, c)$ with $c$ the adiabatic sound speed,
\begin{equation}
\label{eq:sound_speed_sq}
c = \frac{1}{\beta}\, (\Gamma \llangle r_0; r_0 \rrangle)^{-1/2}\, \llangle e_0 - \mu\,\rho_0; e_0 - \mu\,\rho_0 \rrangle^{1/2} \,.
\end{equation}

Through the relations~\eqref{eq:Rproperties} the matrices $A$ and $C$ determine $R$ up to a global sign. One obtains
\begin{equation}\label{Rmatrix}
R = \begin{pmatrix} \langle\tilde{\psi}_{-1}\vert \\ \langle\tilde{\psi}_0\vert \\ \langle\tilde{\psi}_{1}\vert \end{pmatrix}\,,
\end{equation}
where the notation $\langle\,\cdot\,\vert$ denotes a row vector, and
\begin{equation}
\begin{split}
\tilde{\psi}_0 &= \llangle e_0 - \mu\,\rho_0; e_0 - \mu\,\rho_0) \rrangle^{-1/2} \begin{pmatrix} -\mu \\ 0 \\ 1 \end{pmatrix} \,,\\
\tilde{\psi}_{\sigma} &= \llangle e_0 - \mu\,\rho_0; e_0 - \mu\,\rho_0 \rrangle^{-1/2} \, (2\,\Gamma)^{-1/2} \begin{pmatrix} \llangle e_0 - \mu\,\rho_0; e_0 \rrangle \\ 0 \\ -\llangle \rho_0; e_0 - \mu\,\rho_0 \rrangle \end{pmatrix} + (2\, \llangle r_0; r_0) \rrangle)^{-1/2} \begin{pmatrix} 0 \\ \sigma \\ 0 \end{pmatrix} 
\end{split}
\end{equation}
for $\sigma = \pm 1$. The corresponding inverse matrix, containing the eigenvectors of $A$ as columns, is given by
\begin{equation}
R^{-1} = \big( \vert \psi_{-1}\rangle\, \vert \psi_0\rangle\, \vert \psi_{1}\rangle \big)
\end{equation}
with
\begin{equation}
\begin{split}
\psi_0 &= \llangle e_0 - \mu\,\rho_0; e_0 - \mu\,\rho_0 \rrangle^{-1/2} \begin{pmatrix} \llangle \rho_0; e_0 - \mu\,\rho_0 \rrangle \\ 0 \\ \llangle e_0 - \mu\,\rho_0; e_0 \rrangle \end{pmatrix} \,,\\
\psi_{\sigma} &= \llangle e_0 - \mu\,\rho_0; e_0 - \mu\,\rho_0 \rrangle^{-1/2} \, \big(\tfrac{1}{2}\, \Gamma\big)^{1/2} \begin{pmatrix} 1 \\ 0 \\ \mu \end{pmatrix} + \big(\tfrac{1}{2}\, \llangle r_0; r_0 \rrangle\big)^{1/2} \begin{pmatrix} 0 \\ \sigma \\ 0 \end{pmatrix} \,.
\end{split}
\end{equation}

Finally, the $G$ coupling matrices can be determined by inserting $R$ into Eq.~\eqref{eq:Gformula}. First, we calculate the entries of $R\, \mathcal{H}^{\rho} R^{\mathrm{T}}$,
\begin{equation}
\begin{split}
\langle \tilde{\psi}_0 \vert \mathcal{H}^{\rho} \vert \tilde{\psi}_0 \rangle &= 0 \,,\\
\langle \tilde{\psi}_0 \vert \mathcal{H}^{\rho} \vert \tilde{\psi}_{\sigma} \rangle &= \beta^{-2}\, \big(2\, \llangle e_0 - \mu\,\rho_0; e_0 - \mu\,\rho_0 \rrangle\, \llangle r_0; r_0 \rrangle\big)^{-1/2}\, \sigma \,,\\
\langle \tilde{\psi}_{\sigma} \vert \mathcal{H}^{\rho} \vert \tilde{\psi}_{\sigma'} \rangle &= - \beta^{-2}\, \big(\llangle e_0 - \mu\,\rho_0; e_0 - \mu\,\rho_0 \rrangle\, \llangle r_0; r_0 \rrangle\, \Gamma \big)^{-1/2}\, \llangle \rho_0; e_0 - \mu\,\rho_0 \rrangle\, \delta_{\sigma \sigma'}\, \sigma
\end{split}
\end{equation}
for $\sigma, \sigma' = \pm 1$. The entries of $R\, \mathcal{H}^{\mathsf{r}} R^{\mathrm{T}}$ are
\begin{equation}
\begin{split}
\langle \tilde{\psi}_0 \vert \mathcal{H}^{\mathsf{r}} \vert \tilde{\psi}_0 \rangle &= 0 \,,\\
\langle \tilde{\psi}_0 \vert \mathcal{H}^{\mathsf{r}} \vert \tilde{\psi}_{\sigma} \rangle &= \beta^{-2}\, (2\,\Gamma)^{-1/2} \,,\\
\langle \tilde{\psi}_{\sigma} \vert \mathcal{H}^{\mathsf{r}} \vert \tilde{\psi}_{\sigma'} \rangle &= - \beta^{-2}\, \Gamma^{-1} \, \llangle \rho_0; e_0 - \mu\,\rho_0 \rrangle
\end{split}
\end{equation}
and the entries of $R\, \mathcal{H}^{\mathsf{e}} R^{\mathrm{T}}$ are
\begin{equation}
\begin{split}
\langle \tilde{\psi}_0 \vert \mathcal{H}^{\mathsf{e}} \vert \tilde{\psi}_0 \rangle &= 0 \,,\\
\langle \tilde{\psi}_0 \vert \mathcal{H}^{\mathsf{e}} \vert \tilde{\psi}_{\sigma} \rangle &= \beta^{-2}\, \big(2\, \llangle e_0 - \mu\,\rho_0; e_0 - \mu\,\rho_0 \rrangle\, \llangle r_0; r_0 \rrangle\big)^{-1/2}\, \mu\, \sigma \,,\\
\langle \tilde{\psi}_{\sigma} \vert \mathcal{H}^{\mathsf{e}} \vert \tilde{\psi}_{\sigma'} \rangle &= \beta^{-2}\, \big(\llangle e_0 - \mu\,\rho_0; e_0 - \mu\,\rho_0 \rrangle\, \llangle r_0; r_0 \rrangle\, \Gamma \big)^{-1/2}\, \llangle e_0 - 2\,\mu\,\rho_0; e_0 - \mu\,\rho_0 \rrangle\, \delta_{\sigma \sigma'}\, \sigma \,.
\end{split}
\end{equation}

Using Eq.~\eqref{eq:Gformula} and \eqref{eq:sound_speed_sq}, one obtains for the coupling matrices 
\begin{equation}
G^0 = \frac{c}{2 \beta}\, \llangle e_0 - \mu\,\rho_0; e_0 - \mu\,\rho_0 \rrangle^{-1/2} \begin{pmatrix} -1 & 0 & 0 \\ 0 & 0 & 0 \\ 0 & 0 & 1 \end{pmatrix}
\end{equation}
and
\begin{equation}
\begin{split}
G^1 &= \frac{c}{2 \beta}\, \llangle e_0 - \mu\,\rho_0; e_0 - \mu\,\rho_0 \rrangle^{-1/2} \left(\Upsilon \begin{pmatrix} -1 & 0 & 1 \\ 0 & 0 & 0 \\ 1 & 0 & 3 \end{pmatrix} + \begin{pmatrix} 0 & 0 & 0 \\ 0 & 0 & 1 \\ 0 & 1 & 0 \end{pmatrix} \right)\,,
\end{split}
\end{equation}
where
\begin{equation}
\label{eq:upsilon_def}
\Upsilon = - \llangle \rho_0; e_0 - \mu\,\rho_0 \rrangle (2 \Gamma)^{-1/2}\,.
\end{equation}
The matrix $G^{-1}$ is determined by $G^{-1} = -(G^1)^{\mathcal{T}}$, where ${}^\mathcal{T}$ denotes the transpose relative to the anti-diagonal. As it has to be, the $G$ coupling matrices are symmetric. 

Note that the thermodynamic averages and cumulants appearing in this paragraph can be obtained as appropriate derivatives of the free energy $F(\mu,\nu,\beta)$ with respect to $\mu$, $\nu$ and $\beta$, see Eqs.~\eqref{eq:avr_derivF} and \eqref{eq:davr_cumulant2}. For example, evaluated at $\nu = 0$
\begin{align}
\llangle e_0 - \mu\,\rho_0; e_0 - \mu\,\rho_0 \rrangle &= - \partial_{\beta}^2\, (\beta F) \,,\\
\llangle \rho_0; e_0 - \mu\,\rho_0\rrangle &= \partial_{\beta} \partial_{\mu} F\,,\\
\llangle \rho_0; \rho_0 \rrangle \llangle e_0; e_0 \rrangle - \llangle \rho_0; e_0 \rrangle^2
&= \beta^{-1}\, \partial_{\beta}^2(\beta F)\,\partial_\mu^2 F - (\partial_{\beta} \partial_{\mu} F)^2\,.
\end{align}
Also expressions involving $r_0$, as for example $\llangle r_0; r_0 \rrangle$, can be written as derivative of $F$ with respect to $\nu$ evaluated at $\nu = 0$. The numerical computation of $F(\mu,\nu,\beta)$ will be described below.

We have arrived at two important qualitative results. Firstly,  $G_{00}^0 = 0$ always and, secondly  $G^1_{11} > 0$, at least for low temperatures, as to be discussed below, and most likely in the regime characterized by Eq.~\eqref{eq:lt_parameter_regime}.

\paragraph{Asymptotic scaling.}
We define the low temperature correlator by
\begin{equation}
S_\mathrm{lt}(j,t) = \big\langle \lvert \rho_{j}(t),r_j(t),e_j(t)\rangle;\langle \rho_{0}(0),r_0(0),e_0(0)\rvert \big\rangle_{\mu,\beta}
\end{equation}
as a $3 \times 3$ matrix, denoting by $\lvert\cdot\rangle$ the column vector and by $\langle \cdot \rvert$ its transpose. The average is with respect to the thermal state \eqref{eq:ensembleDNLS_lt} at $\nu = 0$ and the dynamics is governed by $H_\mathrm{lt}$, compare with \eqref{eq:motionHlt}. We want to relate $S_\mathrm{lt}$ to the DNLS correlator defined by
\begin{equation}\label{eq:S_def}
S(j,t) = \big\langle \lvert \rho_{j}(t),\tilde{r}_j(t),e_j(t)\rangle;\langle \rho_{0}(0),\tilde{r}_0(0),e_0(0)
\rvert\big\rangle_{\mu,\beta}\,.
\end{equation}
Now the dynamics is generated by the DNLS hamiltonian $H$, compare with \eqref{eq:motionDNLS}. In \eqref{eq:S_rho_e_def} we only considered the correlator for $(\rho_j, e_j)$. But now we include also the phase difference $\tilde{r}_j = \Theta(\varphi_{j+1} - \varphi_j)$. We claim that, under the condition \eqref{eq:lt_parameter_regime}, in approximation
\begin{equation}
S(j,t) \simeq S_\mathrm{lt}(j,t)\,.
\end{equation}

The asymptotics of the correlator on the right can be obtained from nonlinear fluctuating hydrodynamics. We use the matrix $R$, see \eqref{Rmatrix}, to transform to normal modes and define
\begin{equation}
\label{eq:sharp}
S^{\sharp}_{\mathrm{lt}}(j,t) = R\,S_{\mathrm{lt}}(j,t)\,R^\mathrm{T}\,,\qquad S^{\sharp}(j,t) = R\,
S(j,t)\,R^\mathrm{T} \,.
\end{equation}
$S^{\sharp}_{\mathrm{lt}}(j,t)$ is approximately diagonal with matrix elements
\begin{equation}
S^\sharp_{\mathrm{lt},\alpha\alpha'}(j,t) \simeq \delta_{\alpha\alpha'}f_\alpha(j,t)\,.
\end{equation}
The scaling form of $f_{\alpha}(j,t)$ is explained in \cite{SpohnAHC2014, MeSp13}. For the sound peak one finds
\begin{equation}\label{eq:sound_peak_scaling}
f_\sigma(x,t)\simeq (\lambda_\mathrm{s} t)^{-2/3} f_{\mathrm{KPZ}} \big((\lambda_\mathrm{s} t)^{-2/3}(x- \sigma ct)\big)\,,
\end{equation}
$\sigma = \pm 1$, with the non-universal scaling coefficient
\begin{equation}
\lambda_\mathrm{s} = 2 \sqrt{2}\, \lvert G^\sigma_{\sigma\sigma} \rvert \,.
\end{equation}
The universal scaling function $f_{\mathrm{KPZ}}$ is tabulated in \cite{PrSp2004}, denoted there by $f$. $f_{\mathrm{KPZ}}\geq 0$, $\int \ud x \, f_{\mathrm{KPZ}}(x) = 1$, $f_{\mathrm{KPZ}}(x) = f_{\mathrm{KPZ}}(-x)$, $\int \ud x\, x^2\, f_{\mathrm{KPZ}}(x) \simeq 0.510523$. $f_{\mathrm{KPZ}}$ looks like a Gaussian with a large $\lvert x \rvert$ decay as $\exp[-0.295 \lvert x \rvert^{3}]$. Plots are provided in~\cite{PrSp2000, PrSp2004}. The shape function for the heat peak is more easily written in Fourier space,
\begin{equation}\label{eq:heat_peak_scaling}
\hat{f}_0(k,t)\simeq \mathrm{e}^{-\lvert k \rvert^{5/3} \lambda_\mathrm{h} t}\,,
\end{equation}
with the non-universal coefficient
\begin{equation}
\lambda_\mathrm{h} = \lambda_\mathrm{s}^{-2/3}\, (G^0_{\sigma\sigma})^2\, (4 \pi)^2 \int^\infty_0 \ud t\, t^{-2/3} \,\cos (2 \pi c t) \int_{\mathbb{R}} \ud x\, f_{\mathrm{KPZ}}(x)^2\,. 
\end{equation}

\paragraph{Numerical evaluation of the free energy.}
To compute the partition function
\begin{equation}
Z_N(\mu,\nu,\beta) = \int \mathrm{e}^{-\beta\left(H - \mu \mathsf{N} - \nu \sum_j \tilde{r}_j \right)} \prod_{j=0}^{N-1} \ud\psi_j \ud\psi_j^* \,,
\end{equation}
we first switch to polar coordinates $(\rho_j, \varphi_j)$. For the case $\nu = 0$ we explicitly calculate the $\varphi_j$ integrals \cite{RasmussenPRL2000}, i.e., we integrate with respect to $\tilde{r}_j$ on $[-\pi, \pi]$. This leads to
\begin{equation}
Z_N(\mu,0,\beta) = \int \prod_{j=0}^{N-1} K(\rho_{j+1}, \rho_j) \, \ud\rho_j
\end{equation}
with the \emph{transfer operator} or kernel $K(x, y) = K_1(x, y) K_0(y)$ and
\begin{equation}
\label{eq:transfer_operator}
K_1(x, y) = 2\pi I_0\big(\beta \tfrac{1}{m} \sqrt{x\,y}\big)\, \mathrm{e}^{-\beta \frac{1}{2m} (x + y)} \,, \qquad
K_0(y) = \mathrm{e}^{\beta \frac{1}{2} \mu^2/g}\, \mathrm{e}^{-\beta \frac{1}{2} g \left(y - \frac{\mu}{g}\right)^2}
\,.
\end{equation}
$I_0$ is the modified Bessel function of the first kind. Then
\begin{equation}
\label{eq:lambda_max_limit}
\lim_{N \to \infty} \tfrac{1}{N} \log Z_N(\mu,0,\beta) = \log(\lambda_{\max}(K))\,,
\end{equation}
where $\lambda_{\max}$ denotes the largest eigenvalue. We follow the ideas in \cite{Nystrom1930, Bornemann2010} and use a Nystr\"om-type discretization for the kernel. Given a Gauss quadrature rule as
\begin{equation}
\label{eq:quadrature_rule}
\int_0^{\infty} f(\rho)\,\mathrm{e}^{-\beta \frac{1}{2} g \left(\rho - \frac{\mu}{g}\right)^2} \ud\rho \approx \sum_{i=1}^n w_i\,f(x_i)
\end{equation}
with positive weights $w_i$ and base points $x_i$ \cite{GolubWelsch}, we construct the symmetric matrix
\begin{equation}
\big( K_1(x_i, x_{i'})\,\sqrt{w_i\,w_{i'}} \big)_{i,i'=1}^n
\end{equation}
and calculate its largest eigenvalue, denoted $\lambda_1$. Then
\begin{equation}
\log(\lambda_{\max}(K)) \approx \beta\,\tfrac{1}{2} \tfrac{\mu^2}{g} + \log \lambda_1 \,.
\end{equation}
Numerically, we observe exponential convergence with respect to the number of quadrature points. At $\beta = 15$ and $\mu = 1$ for example, $n = 16$ points suffice for double precision accuracy. For non-zero $\nu$ we proceed analogously. The angular integral is no longer given by a special function and we have to determine it numerically.

Based on \eqref{eq:lambda_max_limit}, one then obtains thermodynamic averages and cumulants as appropriate derivatives of the canonical free energy
\begin{equation}
F(\mu,\nu,\beta) = - \beta^{-1} \lim_{N \to\infty} \tfrac{1}{N} \log Z_{N}(\mu,\nu,\beta)\
\end{equation}
with respect to $\mu$, $\nu$, and $\beta$. For example, we determine $\mu$ numerically such that $\langle \rho_j \rangle = 1$, as summarized in the following table for several values of $\beta$:
\begin{center}
\begin{tabular}{r|cccccccc}
$\beta$ & 1       & 2       & 5       & 10       & 15       & 20       & 100        & 200 \\
\hline
$\mu$   & 1.05627 & 1.18426 & 1.08815 &  1.03863 &  1.02489 &  1.01839 &   1.003562 &   1.001774
\end{tabular}
\end{center}

\begin{figure}[!ht]
\centering
\includegraphics[width=0.4\textwidth]{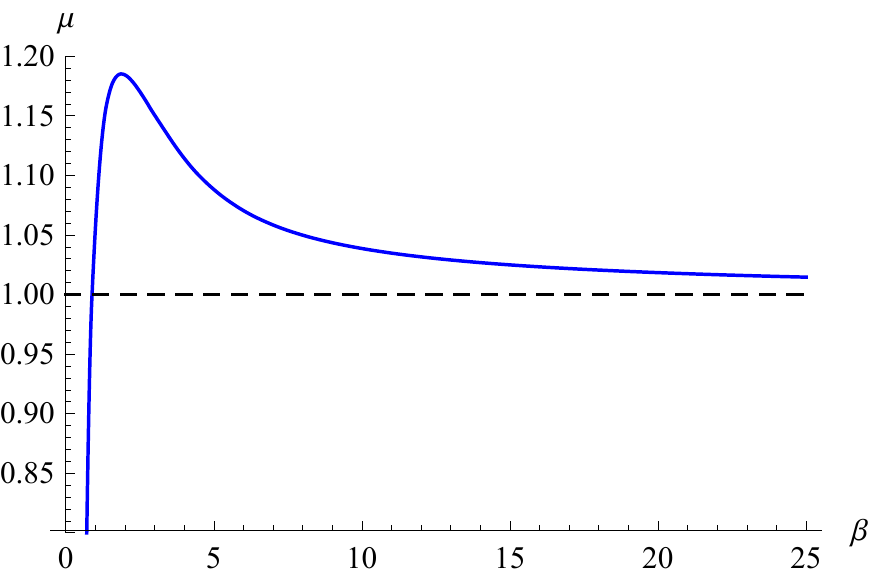}
\caption{The chemical potential $\mu$ as a function of $\beta$ such that $\langle \rho_j \rangle = 1$, for the DNLS with parameters $m = 1$ and $g = 1$.}
\label{fig:nonintegrable_mu_beta}
\end{figure}
Fig.~\ref{fig:nonintegrable_mu_beta} visualizes $\mu$ as a function of $\beta$.

\paragraph{Low temperature expansion.}
For the special case $\nu = 0$, we expand the entries of $C$, $A$, $R$, $\vec{G}$ in terms of $\beta^{-1}$ at fixed value of the average density $\bar{\rho}$. For that purpose, we expand the hamiltonian around the ground state by regarding $r_j$ and $z_j = \rho_j - \bar{\rho}$ as small. We switch to the moving frame, which corresponds to replacing $H$ by $\check{H} = H - g \bar{\rho}\,(\mathsf{N} - \bar{\rho} N) - \tfrac{1}{2} g \bar{\rho}^2 N$ with the ground state energy subtracted. Taylor expansion up to third order results in
\begin{equation}
\check{H} = \check{H}_{\mathrm{ex}} + \sum_{j=0}^{N-1} \Big( {-}\tfrac{1}{16 m \bar{\rho}^2} (z_{j+1}+z_j) (z_{j+1}-z_j)^2 + \mathcal{O}(r_j^4) + \mathcal{O}(z_j^2\, r_j^2) + \mathcal{O}(z_j^4) \Big)\,.
\end{equation}
The expansion hamiltonian $\check{H}_{\mathrm{ex}}$ is given by
\begin{equation}
\label{eq:Heff}
\check{H}_{\mathrm{ex}} = \sum_{j=0}^{N-1} \mathfrak{e}_j, \qquad \mathfrak{e}_j = \tfrac{1}{2m} r_j^2\,\big(\bar{\rho} + \tfrac{1}{2}(z_j + z_{j+1})\big) + \tfrac{1}{8 m \bar{\rho}}(z_{j+1}-z_j)^2 + \check{V}(z_j)
\end{equation}
with potential 
\begin{equation}
\check{V}(x) = \tfrac{1}{2} g\, x^2 \quad \text{for}\quad x > -\bar{\rho}\,,\qquad \check{V}(x) = \infty \quad \text{for}\quad x \leq -\bar{\rho}\,.
\end{equation}
The boundary condition enforces the constraint $z_j \ge -\bar{\rho}$. Note that the expansion does not depend on $\bar{\varphi}$. Clearly $\check{H}_{\mathrm{ex}}$ is stable and we have found a state of minimal energy. The term cubic in $z_j$ is unstable, but will be stabilized by higher order terms in the expansion. These are expected to be small corrections to the quantum pressure $\tfrac{1}{8 m \bar{\rho}}(z_{j+1}-z_j)^2$ and hence will be neglected.

The conserved fields of $\check{H}_{\mathrm{ex}}$ are $z_j$, $r_j$, $\mathfrak{e}_j$. Eventually we want to relate them to the original field variables. $z_j$ transforms to $\rho_j$ since the constant shift by $\bar{\rho}$ drops out from the cumulants below. Concerning the local energy,
\begin{equation}
e_j \simeq \mathfrak{e}_j + g \bar{\rho}\,z_j + \tfrac{1}{2} g \bar{\rho}^2\,, \quad \text{that is}, \quad \mathfrak{e}_j \simeq e_j - g \bar{\rho}\,\rho_j + \mathrm{const} \,,
\end{equation}
such that in a cumulant $\mathfrak{e}_j$ approximates $e_j - \mu\,\rho_j$.

We use $\check{H}_{\mathrm{ex}}$ to evaluate the partition function
\begin{equation}
\check{Z}_N(\check{\mu},\beta) = \int_{\mathbb{R}^{2N}} \mathrm{e}^{-\beta \left( \check{H}_{\mathrm{ex}} - \check{\mu} \sum_{j=0}^{N-1} z_j \right)} \prod_{j=0}^{N-1} \ud z_j\, \ud r_j\,.
\end{equation}
The $r_j$ integrals can be obtained in closed form, and for the $z_j$ integrals we proceed as
\begin{equation}
\int_{\mathbb{R}} \ud z\, \tfrac{\mathrm{e}^{-\beta \frac{1}{2} z^2}}{\sqrt{2\pi / \beta}} \, f(z) \approx \int_{\mathbb{R}} \ud z\, \tfrac{\mathrm{e}^{-\beta \frac{1}{2} z^2}}{\sqrt{2\pi / \beta}} \, \big( f(0) + f'(0)\,z + \tfrac{1}{2} f''(0)\,z^2 + \dots \big) = f(0) + \tfrac{1}{2\beta}\, f''(0) + \dots \,.
\end{equation}
One obtains
\begin{equation}
\langle z_j \rangle = \tfrac{\check{\mu}}{g} - \tfrac{1}{2 g} \big( \tfrac{\check{\mu}}{g} + \bar{\rho} \big)^{-1}\, \beta^{-1} + \mathcal{O}(\beta^{-2}) \,.
\end{equation}
Setting $\langle z_j \rangle = 0$ then yields
\begin{equation}
\label{eq:mu_asymptotic}
\check{\mu} = (2 \bar{\rho}\,\beta)^{-1} + \mathcal{O}(\beta^{-2}) \,.
\end{equation}
To leading order the average energy is given by
\begin{equation}
\langle \mathfrak{e}_j \rangle = \tfrac{\check{\mu}^2}{2 g} + \Big( 1 - \tfrac{\check{\mu}}{2 g} \big( \tfrac{\check{\mu}}{g} + \bar{\rho} \big)^{-1} \Big) \beta^{-1} + \mathcal{O}(\beta^{-2}) \,,
\end{equation}
and inserting \eqref{eq:mu_asymptotic} results in
\begin{equation}
\langle \mathfrak{e}_j \rangle = \beta^{-1} + \mathcal{O}(\beta^{-2}) \quad \text{for } \langle z_j \rangle = 0 \,.
\end{equation}
Using relations analogous to \eqref{eq:davr_cumulant2}, one obtains
\begin{align}
\llangle z_0; z_0 \rrangle &= \tfrac{1}{g}\, \beta^{-1} + \tfrac{1}{2 g^2} \big( \tfrac{\check{\mu}}{g} + \bar{\rho} \big)^{-2}\, \beta^{-2} + \mathcal{O}(\beta^{-3}) \,, \\
\llangle z_0; \mathfrak{e}_0 \rrangle &= \tfrac{\check{\mu}}{g}\, \beta^{-1} - \tfrac{\bar{\rho}}{2 g} \big( \tfrac{\check{\mu}}{g} + \bar{\rho} \big)^{-2}\, \beta^{-2} + \mathcal{O}(\beta^{-3}) \,, \\
\llangle \mathfrak{e}_0; \mathfrak{e}_0 \rrangle &= \tfrac{\check{\mu}^2}{g}\, \beta^{-1} + \tfrac{1}{2} \Big( 1 + \bar{\rho}^2 \big( \tfrac{\check{\mu}}{g} + \bar{\rho} \big)^{-2} \Big) \beta^{-2} + \mathcal{O}(\beta^{-3}) \,,
\end{align}
and inserting \eqref{eq:mu_asymptotic} results in
\begin{align}
\llangle z_0; z_0 \rrangle &= \tfrac{1}{g}\, \beta^{-1} + \tfrac{1}{2} (g \bar{\rho}\,\beta)^{-2} + \mathcal{O}(\beta^{-3}) \,, \\
\llangle z_0; \mathfrak{e}_0 \rrangle &= 0 + \mathcal{O}(\beta^{-3}) \,, \\
\llangle \mathfrak{e}_0; \mathfrak{e}_0 \rrangle &= \beta^{-2} + \mathcal{O}(\beta^{-3})
\end{align}
for $\langle z_j \rangle = 0$. It follows that
\begin{equation}
\llangle \mathfrak{e}_0 - \check{\mu}\,z_0; \mathfrak{e}_0 - \check{\mu}\,z_0 \rrangle = \beta^{-2} + \mathcal{O}(\beta^{-3})
\end{equation}
and
\begin{equation}
\Gamma = \tfrac{1}{g}\, \beta^{-3} + \mathcal{O}(\beta^{-4}) \,.
\end{equation}
The variance of the stretch is
\begin{equation}
\llangle r_0; r_0 \rrangle = m \big( \tfrac{\check{\mu}}{g} + \bar{\rho} \big)^{-1}\,\beta^{-1} + \mathcal{O}(\beta^{-2}) \,, \quad \llangle r_0; r_0 \rrangle = \tfrac{m}{\bar{\rho}} \beta^{-1} + \mathcal{O}(\beta^{-2}) \text{ for } \langle z_j \rangle = 0 \,.
\end{equation}
We have collected all ingredients to calculate the remaining thermodynamic quantities. One obtains for the square of the sound speed
\begin{equation}
c^2 = \tfrac{1}{m} ( \check{\mu} + g\,\bar{\rho}) + \mathcal{O}(\beta^{-1}), \qquad c^2 = \tfrac{1}{m} g\,\bar{\rho} + \mathcal{O}(\beta^{-1}) \quad \text{for } \langle z_j \rangle = 0 \,.
\end{equation}
The coupling matrix $G^0$ is
\begin{equation}
G^0 = \tfrac{1}{2} \sqrt{\tfrac{1}{m} ( \check{\mu} + g\,\bar{\rho})} \begin{pmatrix} -1 & 0 & 0 \\ 0 & 0 & 0 \\ 0 & 0 & 1 \end{pmatrix} + \mathcal{O}(\beta^{-1})
\end{equation}
and the inner term of $G^1$ defined in \eqref{eq:upsilon_def} is
\begin{equation}
\Upsilon = \tfrac{1}{2} \big( \tfrac{\check{\mu}}{g} + \bar{\rho} \big)^{-1} (2 g\,\beta)^{-1/2} + \mathcal{O}(\beta^{-3/2}) \,.
\end{equation}
In particular, $G^1_{11} > 0$ as asserted above.

The square of the matrix elements of $R^{-1}$ are the Landau-Placzek ratios. For the central column of $R^{-1}$ one obtains
\begin{equation}
\begin{split}
\psi_0 &= \big( -\tfrac{1}{2} ( \check{\mu} + g\,\bar{\rho} )^{-1}, 0, 1 - \tfrac{\check{\mu}}{2} ( \check{\mu} + g\,\bar{\rho} )^{-1} \big)^{\mathrm{T}} \beta^{-1} + \mathcal{O}(\beta^{-2}) \,,\\
\psi_0 &= \big( -\tfrac{1}{2 g \bar{\rho}}, 0, 1 \big)^{\mathrm{T}} \beta^{-1} + \mathcal{O}(\beta^{-2}) \quad \text{for } \langle z_j \rangle = 0 \,,
\end{split}
\end{equation}
and for the left and right column
\begin{equation}
\psi_{\sigma} = \Big((2 g\,\beta)^{-1/2} + \mathcal{O}(\beta^{-3/2}) \Big) \begin{pmatrix} 1 \\ 0 \\ \check{\mu} \end{pmatrix} + \Big(\big( \tfrac{\check{\mu}}{g} + \bar{\rho} \big)^{-1/2} (2 \beta / m)^{-1/2} + \mathcal{O}(\beta^{-3/2}) \Big)\begin{pmatrix} 0 \\ \sigma \\ 0 \end{pmatrix} \,,
\end{equation}
\begin{equation}
\psi_{\sigma} = \begin{pmatrix} (2 g\,\beta)^{-1/2} + \mathcal{O}(\beta^{-3/2}) \\[1ex] (2 \bar{\rho}\,\beta/m)^{-1/2}\, \sigma + \mathcal{O}(\beta^{-3/2}) \\[1ex] \tfrac{1}{\bar{\rho}} (8 g)^{-1/2}\,\beta^{-3/2} + \mathcal{O}(\beta^{-5/2}) \end{pmatrix} \quad \text{for } \langle z_j \rangle = 0 \,.
\end{equation}
In particular our results imply that, to leading order in $\beta^{-1}$, the peak weights for the $\rho$-$\rho$ correlations are given by
\begin{equation}
\big( (2 g \beta)^{-1}, (2 g \bar{\rho} \beta)^{-2}, (2 g \beta)^{-1} \big) \,,
\end{equation}
for the $r$-$r$ correlations by 
\begin{equation}
(2 \bar{\rho}\,\beta/m)^{-1} (1, 0, 1)\,,
\end{equation}
and for the $\mathfrak{e}$-$\mathfrak{e}$ correlations by
\begin{equation}
\big( (8 g \bar{\rho}^2)^{-1} \beta^{-3}, \beta^{-2}, (8 g \bar{\rho}^2)^{-1} \beta^{-3} \big) \,.
\end{equation}

In the simulations \cite{KulkarniLamacraft2013} the $\rho$-$\rho$ correlations showed no heat peak, which came unexpected at first glance. From our low temperature expansion one concludes that the heat peak is suppressed by a relative factor of $1/\beta$, $1/\beta = 0.005$ in \cite{KulkarniLamacraft2013}.

\section{Molecular dynamics simulations}
\label{sec6}
\paragraph{Time evolution.}
To solve the DNLS differential equation \eqref{eq:nonintNLS} numerically, the symplectic fourth-order symmetric Runge-Kutta-Nystr\"om method $\mathrm{SRKN}^b_6$ by Blanes and Moan \cite{BlanesMoan2002} is found to be very useful, see also \cite{HairerLubichWanner2006}. It can be regarded as generalization of the Strang splitting technique. Specifically, a canonical approach for solving the DNLS is to split the hamiltonian \eqref{eq:nonintNLS_hamiltonian} into a kinetic and nonlinear part, $H = T + U$, with
\begin{equation}
T = \sum_{j=0}^{N-1} \tfrac{1}{2m} \lvert\psi_{j+1} - \psi_{j}\rvert^2 \,, \qquad U = \sum_{j=0}^{N-1} \tfrac{1}{2}\,g\,|\psi_j|^4 \,.
\end{equation}
Then the flow over time $t$ (separately induced by $T$ and $U$) is exactly given by
\begin{align}
\Phi^T_t: \ \hat{\psi}_k &\mapsto \mathrm{e}^{-\mathrm{i}\,(1 - \cos(2 \pi k/N))\,t/m} \, \hat{\psi}_k \,,\\
\Phi^U_t: \ \psi_j &\mapsto \mathrm{e}^{-\mathrm{i}\,g\,\lvert \psi_j \rvert^2\, t} \, \psi_j \,,
\end{align}
where $\hat{\psi}_k = (\mathcal{F} \psi)_k$ denotes the discrete Fourier coefficient $k$ of $(\psi_j)_{j=0,\dots,N-1}$. The symmetric Runge-Kutta-Nystr\"om method approximates a time step $h$ by the composition
\begin{equation}
\Psi_h = \Phi^U_{b_{s+1} h} \circ \Phi^T_{a_s h} \circ \dots \circ \Phi^U_{b_2 h} \circ \Phi^T_{a_1 h} \circ \Phi^U_{b_1 h}
\end{equation}
with appropriate coefficients $a_i$, $b_i$. They satisfy the symmetry conditions $a_{s+1-i} = a_i$ and $b_{s+2-i} = b_i$. In our case the number of stages $s = 6$, and the numerical values of $a_i$ and $b_i$ for $\mathrm{SRKN}^b_6$ are tabulated in \cite{BlanesMoan2002}. In the simulations below the system size (number of lattice sites) is always $N = 4096$, and the time step $h = \frac{1}{4}$.

\paragraph{Thermodynamic equilibration.}
For each simulation run, we draw an initial state from the canonical ensemble
\begin{equation}
Z_N(\mu,\beta)^{-1}\, \mathrm{e}^{-\beta(H - \mu \mathsf{N})} \prod_{j=0}^{N-1} \ud p_j\, \ud q_j
\end{equation}
for a given inverse temperature $\beta$ and chemical potential $\mu$, where $H$ is the DNLS hamiltonian \eqref{eq:nonintNLS_hamiltonian_pq}. To obtain such a state, we discretize the following fictitious overdamped Langevin dynamics \cite{CancesLegollStoltz2007} (also known as biased random-walk)
\begin{equation}
\begin{split}
\ud p_j(\tau) &= - \partial_{p_j} (H - \mu \mathsf{N})\,\ud \tau + \sqrt{\tfrac{2}{\beta}}\,\ud W_{p,j}(\tau) \,, \\
\ud q_j(\tau) &= - \partial_{q_j} (H - \mu \mathsf{N})\,\ud \tau + \sqrt{\tfrac{2}{\beta}}\,\ud W_{q,j}(\tau) \,, \quad j = 0,\dots,N-1
\end{split}
\end{equation}
where $W_{p,j}(\tau)_{\tau \ge 0}$ and $W_{q,j}(\tau)_{\tau \ge 0}$ are standard Wiener processes. Numerically, we use the Euler-Maruyama method with step size $\Delta \tau$,
\begin{equation}
\begin{split}
p_j^{n+1} &= p_j^n - \Delta\tau\,\partial_{p_j} (H - \mu \mathsf{N}) + \sqrt{\tfrac{2 \Delta\tau}{\beta}}\,G_{p,j}^n \,,\\
q_j^{n+1} &= q_j^n - \Delta\tau\,\partial_{q_j} (H - \mu \mathsf{N}) + \sqrt{\tfrac{2 \Delta\tau}{\beta}}\,G_{q,j}^n \,,
\end{split}
\end{equation}
where the $G_{p,j}^n$ and $G_{q,j}^n$ are independent standard Gaussian variables. In our implementation, we set $\Delta \tau = \frac{1}{64}$ and perform $1024$ such steps. We start with a sample drawn from the canonical ensemble of the quadratic hamiltonian $H_2$, which is obtained from $\check{H}_{\mathrm{ex}}$ of \eqref{eq:Heff} by dropping the cubic term $r_j^2(z_j + z_{j+1})$. $H_2$ is diagonalized by Fourier transformation.

\paragraph{Low temperature dynamics.}
Numerically we simulate the original DNLS \eqref{eq:nonintNLS}, with initial states drawn from the canonical ensemble of $H$, and determine $S^{\sharp}(j,t)$, compare with \eqref{eq:S_def} and \eqref{eq:sharp}. We average over $10^6$ simulation runs. Due to the periodic boundary conditions, the sum of the stretches $\sum_{j=0}^{N-1} \tilde{r}_j$ is necessarily zero or a multiple of $2\pi$. At low temperatures there are no umklapp processes, the sum is zero, meaning that with respect to $\tilde{r}_j$ one simulates a microcanonical ensemble. Since in our theory the canonical average is used, we shift the numerical $S_{\tilde{r}\tilde{r}}(j,t)$ a posteriori by the small value $\frac{1}{N} \llangle r_0; r_0 \rrangle$.

Following the proposal in \cite{DasDhar2015}, the violation of the conservation law for $\tilde{r}_j$ could be measured quantitatively by considering
\begin{equation}
\Gamma(t) = \sum_{j=0}^{N-1} S_{\tilde{r} \tilde{r}}(j,t) = \sum_{j=0}^{N-1} \langle \tilde{r}_j(t); \tilde{r}_0(0) \rangle\,.
\end{equation}
At low temperatures, $\tilde{r}_j = r_j$ and $\Gamma(t) = 0$, for practical purposes. However at high temperatures $\Gamma(t)$ increases and reaches its canonical average value for large $t$. A similar consistency check is based on the matrix sum rule
\begin{equation}
\sum_{j=0}^{N-1} S^{\sharp}(j,t) = \mathbbm{1} \,,
\end{equation}
which holds only if the fields are conserved. For the numerical simulations with $\beta = 15$ (see below), the difference is $\simeq 0.03$ with respect to the $3 \times 3$ matrix $2$-norm at all recorded times $t$.

\medskip

\begin{figure}[!ht]
\centering
\includegraphics[width=0.8\textwidth]{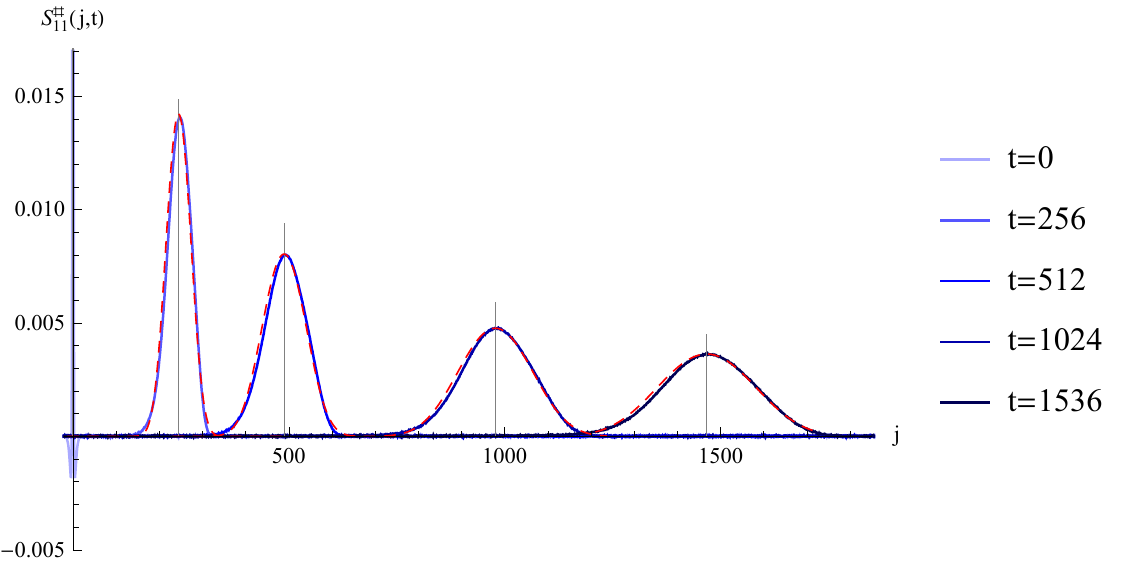}
\caption{Equilibrium two-point correlations $S^{\sharp}_{11}(j,t)$ of the discrete NLS with $m = 1$ and $g = 1$, showing the right-moving sound peak at different time points. Initial states have been drawn from the canonical ensemble \eqref{eq:ensembleDNLS} with parameters $\beta = 15$ and $\mu = 1.02489$. The gray vertical lines show the theoretically predicted sound speed $c$, and the dashed red lines are rescaled KPZ functions.}
\label{fig:canonical_beta15_sound}
\end{figure}
Fig.~\ref{fig:canonical_beta15_sound} visualizes the right-moving sound peak of $S^{\sharp}_{11}(j,t)$ at different time points, with initial states drawn from the canonical ensemble of $H$ with parameters $\beta = 15$ and $\mu = 1.02489$. Following \eqref{eq:sound_peak_scaling}, the red dashed lines are KPZ functions scaled as
\begin{equation}
(\lambda_{\mathrm{s}} t)^{-2/3} f_{\mathrm{KPZ}}\big((\lambda_{\mathrm{s}} t)^{-2/3} (x - c t)\big)
\end{equation}
with $c = 0.9556$ the theoretical sound speed determined by \eqref{eq:sound_speed_sq}. The nonuniversal coefficients $\lambda_{\mathrm{s}}$ have been fitted to minimize the expression
\begin{equation}
\label{eq:L1dist_KPZ}
\sum_{j=0}^{N-1} \big\lvert S^{\sharp}_{11}(j,t) - (\lambda_{\mathrm{s}} t)^{-2/3} f_{\mathrm{KPZ}}\big((\lambda_{\mathrm{s}} t)^{-2/3} (j - c t)\big)\big\rvert \,.
\end{equation}
At the optimal value of $\lambda_{\mathrm{s}}$ the error in \eqref{eq:L1dist_KPZ} is approximately $0.1$ and the largest term in the sum \eqref{eq:L1dist_KPZ} is approximately $2.5 \times 10^{-4}$ at $t = 1536$, thus confirming the scaling exponent and the shape function. The optimal values are recorded in the following table:
\begin{center}
\begin{tabular}{r|ccccc}
$t$ & $256$ & $512$ & $1024$ & $1536$ & theory \\
\hline
$\lambda_{\mathrm{s}}$ & 0.921 & 1.083 & 1.186 & 1.190 & 0.2846
\end{tabular}
\end{center}
Up to the largest accessible time, the value seems to stabilize. However, the theoretical KPZ scaling prediction
\begin{equation}
\lambda_{\mathrm{s}} = 2 \sqrt{2}\,\big\lvert G^{\sigma}_{\sigma\sigma} \big\rvert \,,
\end{equation}
shown as last entry in the table, still deviates by a factor $4$. Such discrepancy, almost perfect scaling plot and substantial deviation for the non-universal coefficients, has been noted before. We refer to \cite{HS15} for a discussion.

\begin{figure}[!ht]
\centering
\subfloat[$t = 256$]{
\includegraphics[width=0.24\textwidth]{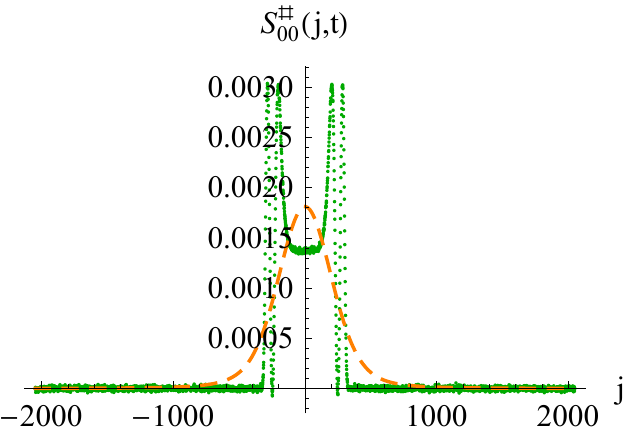}}
\subfloat[$t = 512$]{
\includegraphics[width=0.24\textwidth]{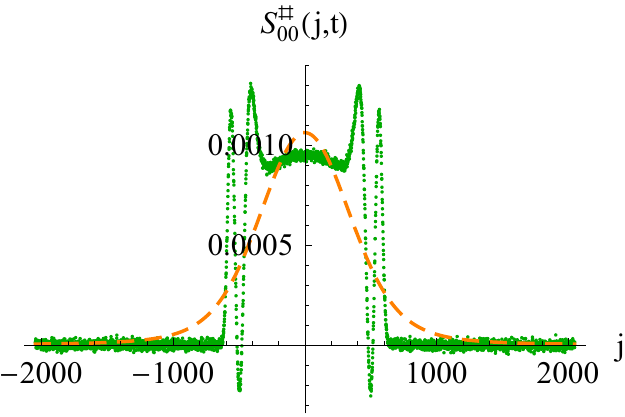}}
\subfloat[$t = 1024$]{
\includegraphics[width=0.24\textwidth]{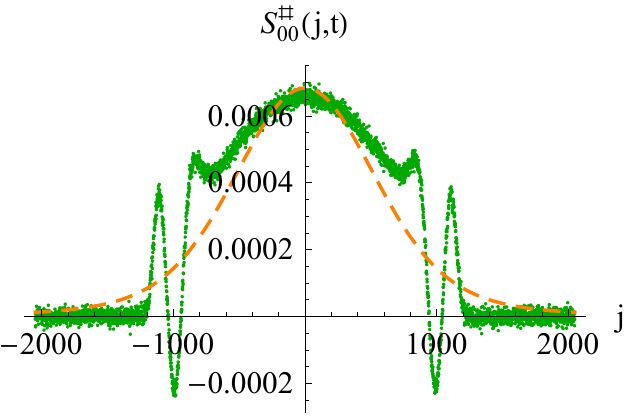}}
\subfloat[$t = 1536$]{
\includegraphics[width=0.24\textwidth]{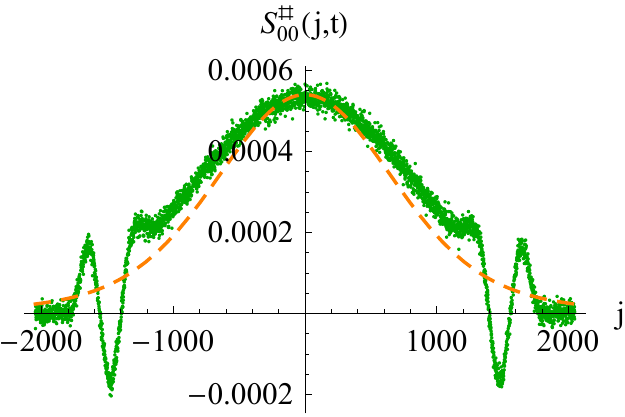}}
\caption{Heat mode $S^{\sharp}_{00}(j,t)$ of the discrete NLS, for the same numerical simulations as in Fig.~\ref{fig:canonical_beta15_sound} at $\beta = 15$. The dashed orange curves show the fitted L\'evy distribution in \eqref{eq:fLevy}.}
\label{fig:canonical_beta15_heat}
\end{figure}
The (relatively broad) heat mode for the same simulations is visualized in Fig.~\ref{fig:canonical_beta15_heat}. One notices some feedback from the sound modes, which seems to weaken as time progresses. According to \eqref{eq:heat_peak_scaling}, mode coupling predicts
\begin{equation}
\label{eq:fLevy}
S^{\sharp}_{00}(j,t) = (\lambda_{\mathrm{h}} t)^{-3/5} f_{{\mathrm{L},5/3}}((\lambda_{\mathrm{h}} t)^{-3/5} j)
\end{equation}
with $f_{\mathrm{L},\alpha}$ the symmetric $\alpha$-stable distribution, also known as $\alpha$-L\'evy distribution, and shown as dashed orange lines in Fig.~\ref{fig:canonical_beta15_heat}. The coefficients $\lambda_{\mathrm{h}}$ are obtained by minimizing
\begin{equation}
\label{eq:L1dist_heat}
\sum_{j=0}^{N-1} \big\lvert S^{\sharp}_{00}(j,t) - (\lambda_{\mathrm{h}} t)^{-3/5} f_{\mathrm{L},5/3}\big((\lambda_{\mathrm{h}} t)^{-3/5} j\big)\big\rvert \,,
\end{equation}
with the result
\begin{center}
\begin{tabular}{r|cccc}
$t$ & $256$ & $512$ & $1024$ & $1536$ \\
\hline
$\lambda_{\mathrm{h}}$ & 17.8 & 21.7 & 22.7 & 22.4
\end{tabular}
\end{center}

\begin{figure}[!ht]
\centering
\subfloat[$t = 256$]{
\includegraphics[width=0.24\textwidth]{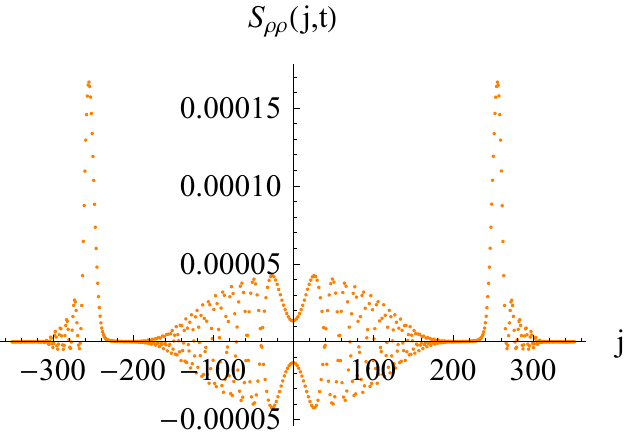}}
\subfloat[$t = 512$]{
\includegraphics[width=0.24\textwidth]{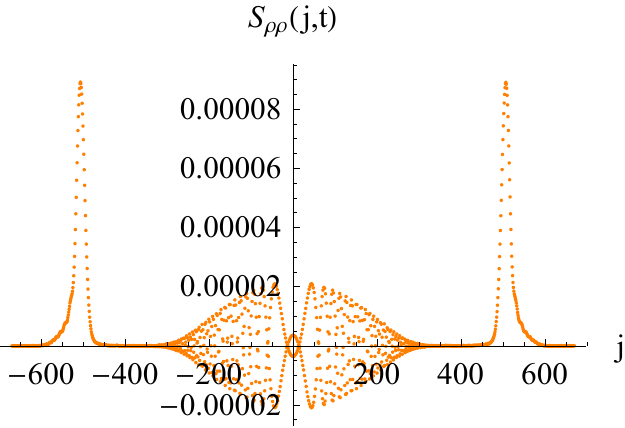}}
\subfloat[$t = 1024$]{
\includegraphics[width=0.24\textwidth]{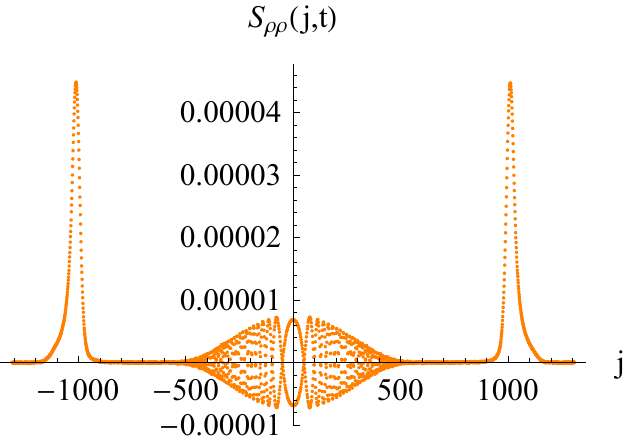}}
\subfloat[$t = 1536$]{
\includegraphics[width=0.24\textwidth]{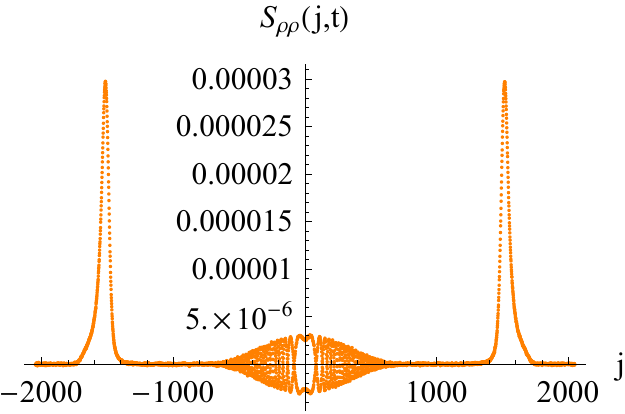}} \\
\subfloat[$t = 256$]{
\includegraphics[width=0.24\textwidth]{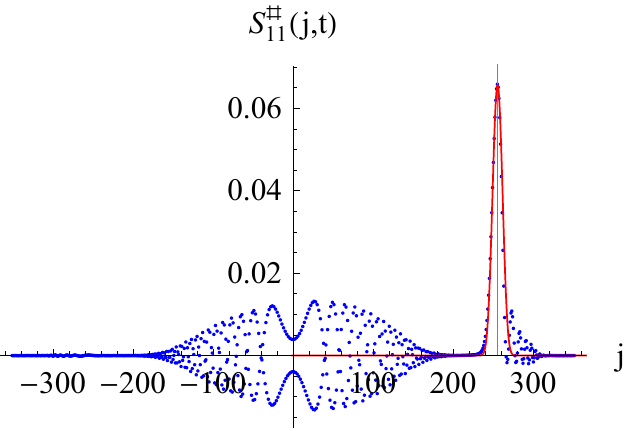}}
\subfloat[$t = 512$]{
\includegraphics[width=0.24\textwidth]{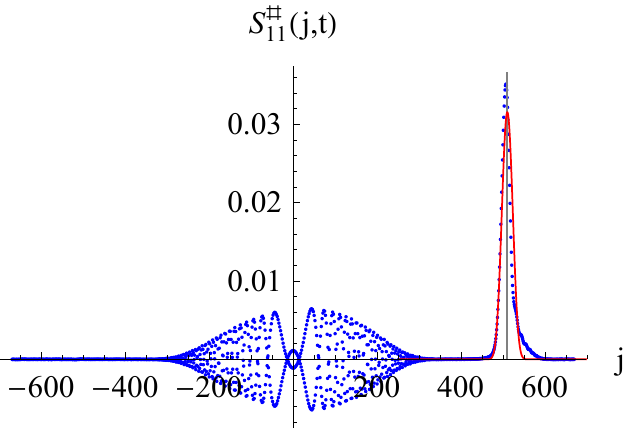}}
\subfloat[$t = 1024$]{
\includegraphics[width=0.24\textwidth]{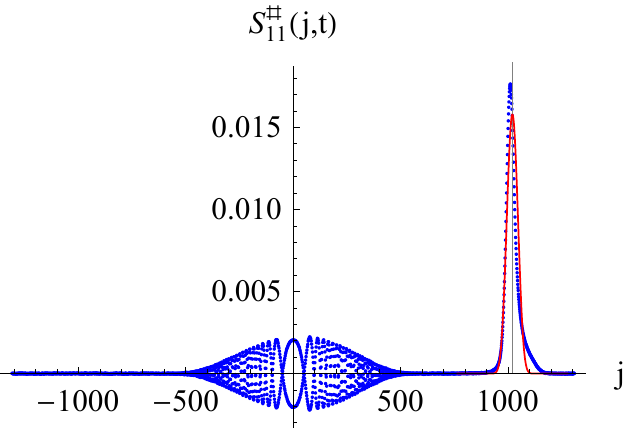}}
\subfloat[$t = 1536$]{
\includegraphics[width=0.24\textwidth]{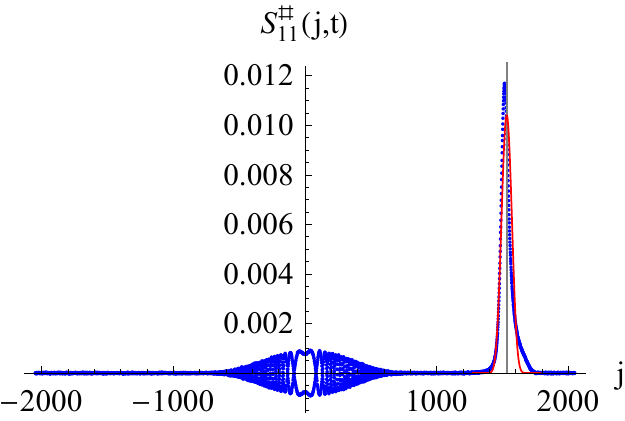}} \\
\subfloat[$t = 256$]{
\includegraphics[width=0.24\textwidth]{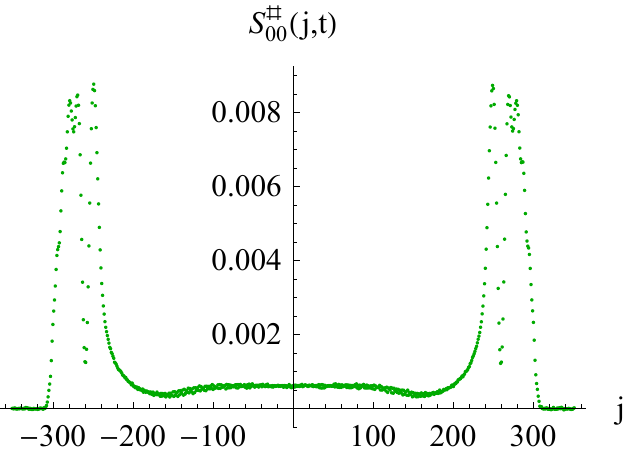}}
\subfloat[$t = 512$]{
\includegraphics[width=0.24\textwidth]{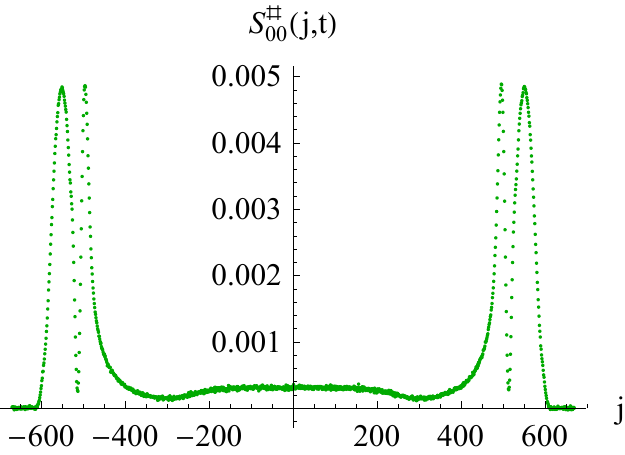}}
\subfloat[$t = 1024$]{
\includegraphics[width=0.24\textwidth]{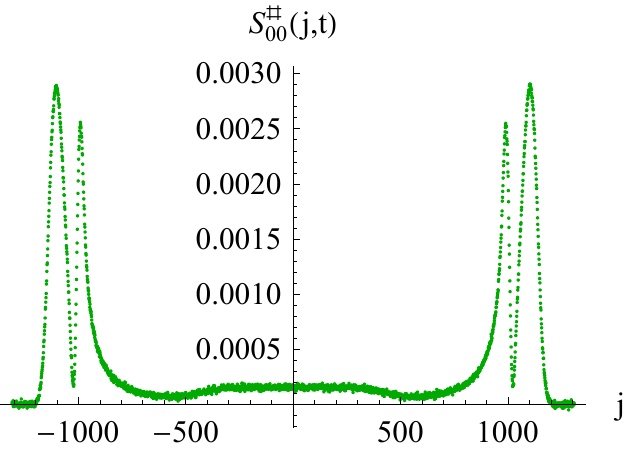}}
\subfloat[$t = 1536$]{
\includegraphics[width=0.24\textwidth]{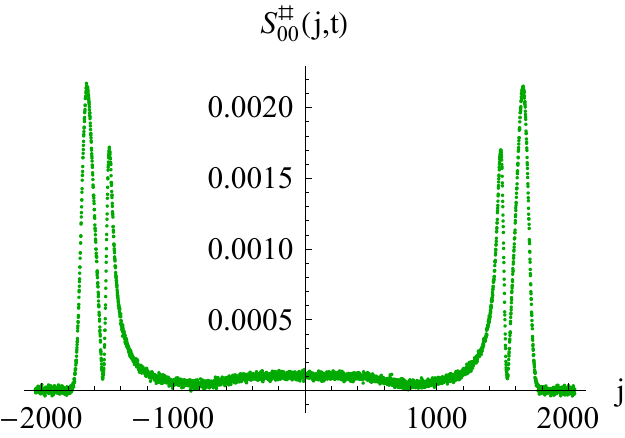}}
\caption{Equilibrium density correlations $S_{\rho\rho}(j,t)$ (top row), sound mode $S^{\sharp}_{11}(j,t)$ (middle row) and heat mode $S^{\sharp}_{00}(j,t)$ (bottom row) of the DNLS at canonical ensemble parameters $\beta = 200$ and $\mu = 1.001774$.}
\label{fig:canonical_beta200}
\end{figure}
To probe the domain of validity for nonlinear fluctuating hydrodynamics we move to even lower temperatures and set $\beta = 200$, still at $\langle \rho_j \rangle = 1$, which are the same parameters as in \cite{KulkarniLamacraft2013}. In the top row of Fig.~\ref{fig:canonical_beta200} we display the density-density correlation. The central peak has dissolved into a bump with rapid oscillations and weakly subballistic spreading. The actual heat peak would have a relative amplitude smaller by a factor of $0.005$ and is thus not visible. The two sound peaks are well separated from the central bump. In \cite{KulkarniLamacraft2013} their structure has been studied in detail for sizes roughly double compared to the one used for Fig.~\ref{fig:canonical_beta200}. There, $k$, $\omega$ space is used, which obstructs a direct comparison. But the numerical evidence \cite{Kulkarni2015} points towards quantitative validity of KPZ scaling. Apparently each sound peak is still governed by a scalar noisy Burgers equation. The energy-energy correlations look very similar to the one in the top row of Fig.~\ref{fig:canonical_beta200}.

Also other linear combinations of the conserved fields can be considered. In the middle row of Fig.~\ref{fig:canonical_beta200} we display $S_{11}^\sharp$. As expected only the right sound peak is left, but the central bump remains as before. The bottom row of Fig.~\ref{fig:canonical_beta200} visualizes $S_{00}^\sharp$. Now the central bump with rapid oscillations has disappeared, giving space for a structure which might be the remnant of the heat peak. On the other side, peaks like the sound peak pop up. Apparently, we see structures which are not accounted for by the hydrodynamic fluctuation theory.

At such ultra-low temperatures, the dynamics is no longer sufficiently chaotic, as implicitly assumed when deriving fluctuating hydrodynamics. For a more precise dynamical underpinning the obvious candidate is the harmonic approximation with hamiltonian $H_2$, compare with last paragraph of Sect.~\ref{sec5}. In MD simulations of the dynamics for $H_2$ at the same parameters, one also observes structures of a similar form as the central bump. A direct comparison seems to be difficult. The best studied example of chain dynamics at ultra-low temperatures is the FPU $\beta$-chain, size order $10^3$ and times up to $10^6$, however at a temperature still a factor $10^{-1}$ lower than studied here \cite{Benettin2013}. For the FPU $\beta$-chain the harmonic approximation is only a small part of the full story, which cautions us to move to rapid conjectures. As a general rule, for integrable systems one is left with a case by case study. To provide an additional piece of evidence we study an integrable version of DNLS in the following section.

\section{The integrable Ablowitz-Ladik model}

The Ablowitz-Ladik (AL) model \cite{AL1976summary, Ablowitz2004} is an \emph{integrable} version of the lattice NLS:
\begin{equation}
\label{eq:ALmodel}
\mathrm{i} \, \tfrac{\ud}{\ud t} \psi_j = -\tfrac{1}{2 m} \Delta \psi_j + \tfrac{1}{2}\, g\,\lvert\psi_j\rvert^2 \left(\psi_{j+1} + \psi_{j-1}\right) \,.
\end{equation}
It is well known \cite[section 3.4]{Ablowitz2004} that \eqref{eq:ALmodel} has a hamiltonian structure with non-canonical Poisson brackets. Specifically, the hamiltonian in our variables is
\begin{equation}
\label{eq:Hal}
H^{\mathrm{al}} = \sum_{j \in \mathbb{Z}} e^{\mathrm{al}}_j \quad \text{with} \quad e^{\mathrm{al}}_j = -\tfrac{1}{2m} (u_{j+1}\,u_j + v_{j+1}\,v_j) - \tfrac{1}{g\,m^2} \log\!\big( 1 - \tfrac{1}{2}\,g m (u_j^2 + v_j^2)\big)
\end{equation}
where we split the wavefunction into its real and imaginary part as
\begin{equation}
\psi_j = \tfrac{1}{\sqrt{2}} (u_j + \mathrm{i} v_j)\,.
\end{equation}
Note that the logarithmic term in \eqref{eq:Hal} requires $\frac{1}{2}\,g m (u_j^2 + v_j^2) = g m \lvert \psi_j \rvert^2 < 1$. The system~\eqref{eq:ALmodel} can then be written as
\begin{equation}
\tfrac{\ud}{\ud t} u_j = c_j\, \partial_{v_j} H^{\mathrm{al}}, \quad \tfrac{\ud}{\ud t} v_j = - c_j\, \partial_{u_j} H^{\mathrm{al}}
\end{equation}
with
\begin{equation}
c_j = 1 - \tfrac{1}{2}\,g m (u_j^2 + v_j^2) \,.
\end{equation}

Due to integrability, the Ablowitz-Ladik model has an infinite number of conservation laws. We just mention the density
\begin{equation}
\rho^{\mathrm{al}}_j = \tfrac{1}{2} \big(\psi_{j+1}\,\psi_j^* + \psi_{j+1}^*\,\psi_j \big) = \tfrac{1}{2} (u_{j+1}\,u_j + v_{j+1}\,v_j)
\end{equation}
with its corresponding current
\begin{equation}
\mathcal{J}^{\mathrm{al}}_{\rho,j} = \tfrac{\mathrm{i}}{2} \big( \tfrac{1}{2m} - \tfrac{1}{2} g \lvert\psi_j\rvert^2 \big) \big( \psi_{j-1}\,\psi_{j+1}^* - \psi_{j-1}^*\,\psi_{j+1} \big),
\end{equation}
for which the following conservation law holds,
\begin{equation}
\tfrac{\ud}{\ud t} \rho^{\mathrm{al}}_j + \mathcal{J}^{\mathrm{al}}_{\rho,j+1} - \mathcal{J}^{\mathrm{al}}_{\rho,j} = 0.
\end{equation}
The local quantity
\begin{equation}
w_j^{\mathrm{al}} = \log(c_j) = \log\big(1 - g m \, \lvert\psi_j\rvert^2 \big)
\end{equation}
is also conserved, with current
\begin{equation}
\mathcal{J}^{\mathrm{al}}_{w,j} = \tfrac{\mathrm{i}}{2}\, g \left(\psi_{j-1}^*\, \partial \psi _{j-1} - \psi_{j-1}\, \partial \psi_{j-1}^* \right).
\end{equation}
Energy is locally conserved with energy current
\begin{equation}
\mathcal{J}^{\mathrm{al}}_{e,j} = -\tfrac{1}{m} \big( \mathcal{J}^{\mathrm{al}}_{\rho,j} + \tfrac{1}{g m} \mathcal{J}^{\mathrm{al}}_{w,j} \big).
\end{equation}

\paragraph{Canonical variables and numerical procedure.}
We apply the ideas in \cite{HairerLubichWanner2006,Tang2007} to derive a hamiltonian structure with canonical Poisson brackets. First, one can absorb the central oscillation by defining, compare with \cite{Tang2007},
\begin{equation}
\chi_j(t) = \mathrm{e}^{\mathrm{i} t / m} \psi_j(t),
\end{equation}
which transforms \eqref{eq:ALmodel} into
\begin{equation}
\label{eq:ALmodel_mod}
\mathrm{i} \, \tfrac{\ud}{\ud t} \chi_j = \big( {-\tfrac{1}{2 m}} + \tfrac{1}{2}\,g\,\lvert\chi_j\rvert^2 \big) \left(\chi_{j+1} + \chi_{j-1}\right).
\end{equation}
We define
\begin{equation}
\sigma(x) = \left(-\frac{\log(1-x)}{x}\right)^{1/2} \quad \text{for} \quad x \in (-\infty,1)
\end{equation}
and set
\begin{equation}
\label{eq:chi2pq}
\begin{split}
q_j &= \sqrt{2}\,\mathrm{Re}(\chi_j)\,\sigma\big(g m\,\lvert\chi_j\rvert^2\big), \\
p_j &= \sqrt{2}\,\mathrm{Im}(\chi_j)\,\sigma\big(g m\,\lvert\chi_j\rvert^2\big).
\end{split}
\end{equation}
Then \eqref{eq:ALmodel_mod} is equivalent to the hamiltonian system
\begin{equation}
\label{eq:AL_canonical}
\tfrac{\ud}{\ud t} p_j = - \partial_{q_j} \tilde{H}^{\mathrm{al}}, \quad \tfrac{\ud}{\ud t} q_j = \partial_{p_j} \tilde{H}^{\mathrm{al}}
\end{equation}
with the hamiltonian
\begin{equation}
\tilde{H}^{\mathrm{al}} = \sum_{j \in \mathbb{Z}} - \tfrac{1}{2m}\,\tau\!\left(\tfrac{1}{2}\,g m \big(p_{j+1}^2 + q_{j+1}^2\big)\right) \tau\!\left(\tfrac{1}{2}\,g m \big(p_j^2 + q_j^2\big)\right) \left(p_{j+1}\,p_j + q_{j+1}\,q_j\right)
\end{equation}
and
\begin{equation}
\tau(x) = \left(\frac{1-\mathrm{e}^{-x}}{x} \right)^{1/2} = 1 - \tfrac{1}{4}\,x + \mathcal{O}(x^2) \quad \text{for} \quad x \in \mathbb{R}.
\end{equation}
Note that $\tilde{H}^{\mathrm{al}}$ is symmetric under $p_j \leftrightarrow q_j$.
The inverse transform to \eqref{eq:chi2pq} is
\begin{equation}
\label{eq:pq2chi}
\chi_j = \tfrac{1}{\sqrt{2}} \big[ q_j\,\tau\!\left(\tfrac{1}{2}\,g m \big(p_j^2+q_j^2\big)\right) + \mathrm{i}\, p_j\,\tau\!\left(\tfrac{1}{2}\,g m \big(p_j^2+q_j^2\big)\right) \big].
\end{equation}

We apply the St\"ormer-Verlet scheme \cite{HairerLubichWanner2006} to numerically integrate \eqref{eq:AL_canonical}, and use Newton iterations to solve the resulting nonlinear systems of equations. For evaluating correlation functions we transform back to the physical coordinates $\psi_j$. Since the procedure is computationally more demanding compared to the nonintegrable DNLS, we use a smaller system size $N = 1024$ for the Ablowitz-Ladik model, and correspondingly a smaller maximum correlation time $t = 512$. We average over $10^5$ simulation runs.

Different from the DNLS, we equilibrate the system by simulating long Hamiltonian trajectories and calculating correlation functions for consecutive, disjoint time intervals of the trajectory. We set $m = g = 1$, and choose an initial state such that the average density $\langle \rho^{\mathrm{al}}_j \rangle \approx \frac{1}{2}$ in order to satisfy the constraint $g m \lvert \psi_j \rvert^2 < 1$.

\begin{figure}[!ht]
\centering
\subfloat[density $S^{\mathrm{al}}_{\rho \rho}(j,t)$]{
\includegraphics[width=0.33\textwidth]{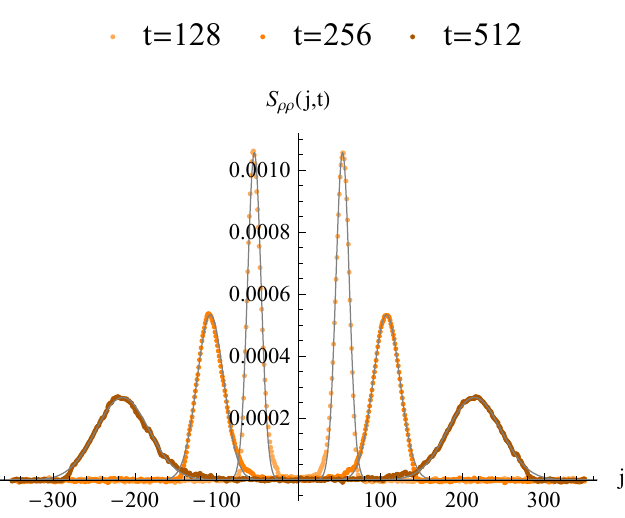}}
\subfloat[stretch $S^{\mathrm{al}}_{r r}(j,t)$]{
\includegraphics[width=0.33\textwidth]{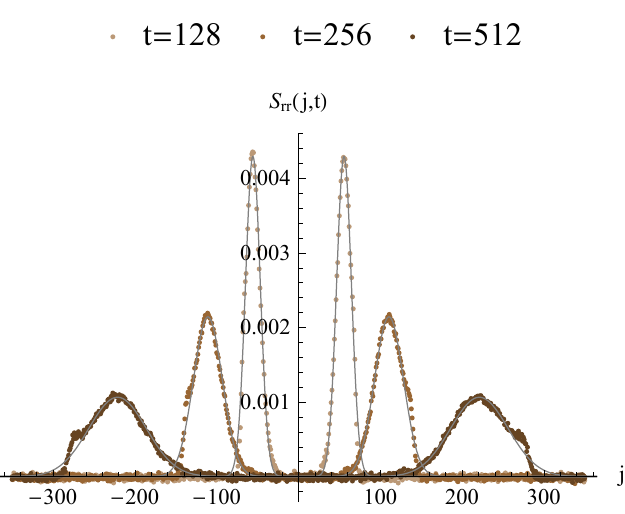}}
\subfloat[energy $S^{\mathrm{al}}_{e e}(j,t)$]{
\includegraphics[width=0.33\textwidth]{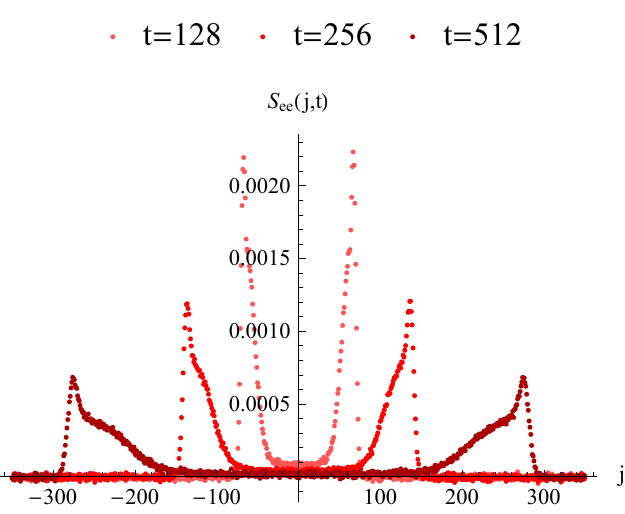}}
\caption{Equilibrium time-correlations of the Ablowitz-Ladik model at inverse temperature $\beta = 15$. Three time points are superimposed. The faint gray curves are fitted Gaussian functions.}
\label{fig:AL_beta15}
\end{figure}

\paragraph{MD simulation results.}
Fig.~\ref{fig:AL_beta15} shows the equilibrium time-correlations of the Ablowitz-Ladik model at inverse temperature $\beta = 15$. The correlations have been shifted by a constant to correct for deviations due to the finite number of samples. As for the nonintegrable DNLS, left and right sound peaks are clearly visible. However, the energy correlations have a sharp cut-off, and the energy peak structure differs from the one of DNLS.

\begin{figure}[!ht]
\centering
\subfloat[density $S^{\mathrm{al}}_{\rho \rho}(j,t)$]{
\includegraphics[width=0.33\textwidth]{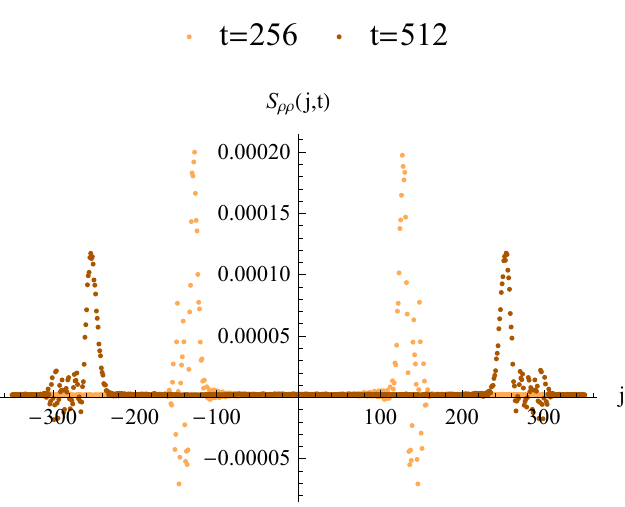}}
\subfloat[stretch $S^{\mathrm{al}}_{r r}(j,t)$]{
\includegraphics[width=0.33\textwidth]{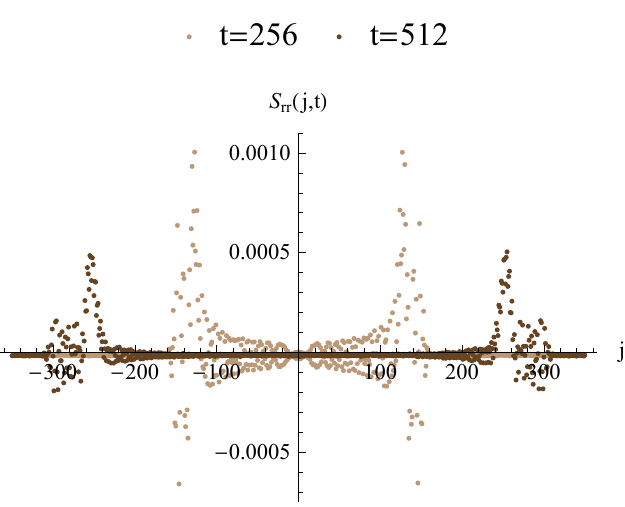}}
\subfloat[energy $S^{\mathrm{al}}_{e e}(j,t)$]{
\includegraphics[width=0.33\textwidth]{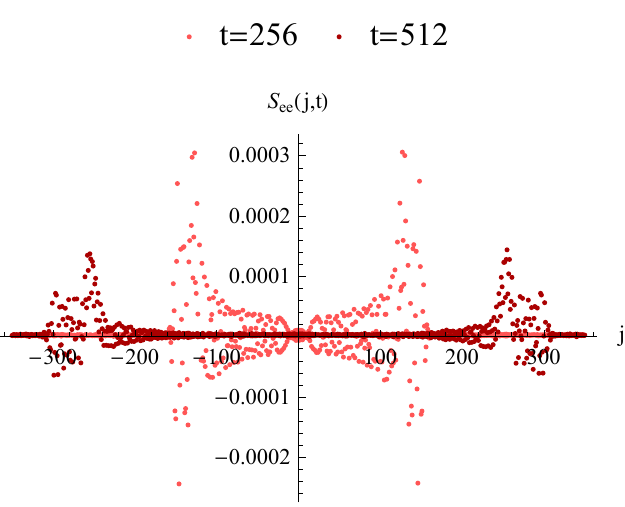}}
\caption{Equilibrium time-correlations of the Ablowitz-Ladik model at inverse temperature $\beta = 200$ and two time points $t = 256$ and $t = 512$ superimposed.}
\label{fig:AL_beta200}
\end{figure}

Due to integrability, one expects ballistic spreading of the correlation functions. As quantitative test, we fit Gaussian functions to the right density peaks by minimizing
\begin{equation}
\label{eq:ALdensityGaussFit}
\sum_{j=0}^{N/2-1} \big\lvert S_{\rho \rho}(j,t) - n_{\rho}\, (\lambda_{\rho} t)^{-1} f_{\mathrm{G}}\big((\lambda_{\rho} t)^{-1} (j - c_{\rho} t)\big)\big\rvert \,.
\end{equation}
Here $n_{\rho}$ is a normalization constant, and the three parameters $\lambda_{\rho}$, $c_{\rho}$, $n_{\rho}$ are optimized numerically, as shown in the following table:
\begin{center}
\begin{tabular}{r|ccc}
$t$ & $128$ & $256$ & $512$ \\
\hline
$\lambda_{\rho}$ & 0.0642 & 0.0639 & 0.0636 \\
$c_{\rho}$ & 0.422 & 0.421 & 0.420
\end{tabular}
\end{center}
The values hardly drift with time, which supports the scaling exponent $-1$ in \eqref{eq:ALdensityGaussFit}.

Fig.~\ref{fig:AL_beta200} shows the very low temperature ($\beta = 200$) equilibrium time-correlations of the Ablowitz-Ladik model, with two time points $t = 256$ and $t = 512$ superimposed. As for the nonintegrable DNLS at $\beta = 200$, the correlations show quickly oscillating features.

\section{Summary and Conclusions}

Our arguments required some length. So let us first summarize what has been accomplished. The DNLS has three dynamical regimes. At density $1$ we explore a single temperature in each regime. At $\beta = 1$, and certainly for smaller $\beta$, the dynamics is characterized by diffusive transport of density and energy. This is the high temperature regime, which is very well confirmed by our numerical simulations. Proceeding towards lower temperatures we subdivide into the regimes of low and ultra-low temperatures.

In the low temperature regime the dynamics is still sufficiently chaotic, but the field of phase differences is conserved, although only with high accuracy. The three conservation laws have equilibrium time correlations whose structure is identical to the one known from generic anharmonic chains. On the theoretical level, the main effort is to obtain the coupling coefficients needed as an input to nonlinear fluctuating hydrodynamics. Without these coefficients, predictions would be mere guess work. Once the couplings are known one can take over the results from \cite{SpohnAHC2014}. The second order expansion of the DNLS Euler currents is more difficult than for anharmonic chains. The reason why we still arrive at fairly explicit results, valid over the entire low temperature regime, relies on the very special structure of the Euler currents, to recall $\langle \mathcal{J}_{\rho}\rangle_{\mu,\nu,\beta} = \nu $, $\langle \mathcal{J}_r\rangle_{\mu,\nu,\beta} = \mu $, $\langle \mathcal{J}_e\rangle_{\mu,\nu,\beta} = \mu\,\nu $. The Euler currents of anharmonic chains have a very different form. For example $\mathcal{J}_r$ would be the momentum which is itself conserved. Still, from the linearized DNLS Euler currents one derives two symmetric sound peaks moving with velocity $\pm c$ and a heat peak standing still. The heat peak broadens as $t^{3/5}$ with a L\'evy $\tfrac{5}{3}$ shape function and the sound peaks broaden as $t^{2/3}$ with KPZ shape function. In the low temperature regime, no matter what the small initial perturbation, and independent of the temperature, one will always obtain a linear combination of these three universal peaks in the long time limit. In our molecular dynamics simulations at $\beta = 15$, the scaling of the sound peaks agrees well with the theory. The heat peak data are still noisy, but show already the characteristic slow spatial decay. For the density-density correlation function the contribution from the heat peak is a factor $\beta^{-1}$ smaller when compared to the sound peak amplitude.

In the ultra-low temperature regime the integrable structure of the dynamics becomes visible. At $\beta = 200$ one observes structures which indicate already substantial contributions from regular motion. Peaks tend to broaden ballistically. The peak structure depends on the particular observable and and on temperature. Universal behavior is lost. As illustrated in the top row of Fig.~\ref{fig:canonical_beta200}, KPZ scaling and integrable correlation structure may coexist. Thus to identify the precise border line of the ultra-low temperature regime seems to be more difficult than distinguishing between the low and high temperature regimes.

The next, but challenging, goal is to numerically obtain comparable low temperature features for the quantized version of the discrete nonlinear Schr\"odinger equation, which is the Bose-Hubbard hamiltonian.

\paragraph{Acknowledgements.} HS thanks Manas Kulkarni and David Huse for stimulating discussions, which started his interest in DNLS. CM thankfully acknowledges computing resources of the Leibniz-Rechenzentrum.

\appendix

\section{Speed of sound, $R$ matrix and $G$ couplings}

We record the theoretical $c$, $R$, and $G$ for the DNLS at our simulation parameters, where we set $m = 1$, $g = 1$. The matrix $G^{-1}$ is specified by the relation $G^{-1} = - (G^{1})^{\mathcal{T}}$, with ${}^\mathcal{T}$ denoting the transpose relative to the anti-diagonal. The entries are rounded to four digits for visual clarity.

At $\beta = 15$ and $\mu = 1.02489$, the speed of sound is $c = 0.9556$ and
\begin{equation}
R =
\begin{pmatrix}
  1.6450 & -2.6311 &  1.0529 \\
-15.0222 &  0      & 14.6575 \\
  1.6450 &  2.6311 &  1.0529 \\
\end{pmatrix}\,, \quad
R^{-1} =
\begin{pmatrix}
 0.1835 & -0.0264 & 0.1835 \\
-0.1900 &  0      & 0.1900 \\
 0.1881 &  0.0412 & 0.1881 \\
\end{pmatrix}\,,
\end{equation}
as well as
\begin{equation}
G^{1} =
\begin{pmatrix}
 -0.0335 & 0      & 0.0335 \\
  0      & 0      & 0.4669 \\
  0.0335 & 0.4669 & 0.1006 \\
\end{pmatrix}\,,
\quad
G^{0} =
\begin{pmatrix}
-0.4669 & 0 & 0      \\
 0      & 0 & 0      \\
 0      & 0 & 0.4669 \\
\end{pmatrix}. \\
\end{equation}

At $\beta = 200$ and $\mu = 1.001774$, the speed of sound is $c = 0.9970$ and
\begin{equation}
R =
\begin{pmatrix}
   6.3731 & -9.9733 &   3.6170 \\
-201.8000 &  0      & 201.4426 \\
   6.3731 &  9.9733 &   3.6170 \\
\end{pmatrix}\,, \quad
R^{-1} = 0.1 \times
\begin{pmatrix}
 0.5002 & -0.0180 & 0.5002 \\
-0.5013 &  0      & 0.5013 \\
 0.5011 &  0.0316 & 0.5011 \\
\end{pmatrix}\,,
\end{equation}
as well as
\begin{equation}
G^{1} =
\begin{pmatrix}
-0.009 & 0      & 0.009  \\
 0     & 0      & 0.5021 \\
 0.009 & 0.5021 & 0.0270 \\
\end{pmatrix}\,,
\quad
G^{0} =
\begin{pmatrix}
-0.5021 & 0 & 0      \\
 0      & 0 & 0      \\
 0      & 0 & 0.5021 \\
\end{pmatrix}. \\
\end{equation}

\section{Average currents}
\label{sec:average_currents}

We establish the relations \eqref{eq:current_avr}. To repeat, the average currents for $H_\mathrm{lt}$ differ from the ones for $H$, the latter having zero average always. Since our expressions for the currents contain derivatives, we use the smooth version of $U$, $V$. Then the Boltzmann factor $\exp[-\beta(H_\mathrm{lt} - \mu \sum_{j=0}^{N-1} \rho_j - \nu \sum_{j=0}^{N-1}r_j)]$ vanishes at $\rho_j = 0$ and $r_j = \pm \pi$. However the Boltzmann factor no longer agrees with the one where $H_\mathrm{lt}$ is replaced by $H$.

For the density current we obtain
\begin{equation}
\begin{split}
\langle\mathcal{J}_{\rho,j}\rangle &= \big\langle \sqrt{\rho_{j-1} \rho_{j}}\,U'(r_{j-1}) \big\rangle \\
&= -\beta^{-1} Z_{\mathrm{lt}}^{-1} \int \big(\partial_{r_{j-1}} \mathrm{e}^{-\beta H_\mathrm{lt}} \big) \big(\mathrm{e}^{\beta\mu \sum_j \rho_j + \beta \nu \sum_j r_j} \big) = \nu\,,
\end{split}
\end{equation}
where we used partial integration in the last step. Correspondingly for the stretch current, 
\begin{equation}
\begin{split}
\langle\mathcal{J}_{r,j}\rangle &= \Big\langle \tfrac{1}{2}\sqrt{\rho_{j+1}/\rho_j}\,U(r_j) + \tfrac{1}{2}\sqrt{\rho_{j-1}/\rho_j}\,U(r_{j-1}) + V'(\rho_j)\Big\rangle \\
&= -\beta^{-1} Z_{\mathrm{lt}}^{-1} \int \big(\partial_{\rho_{j}} \mathrm{e}^{-\beta H_\mathrm{lt}} \big) \big( \mathrm{e}^{\beta\mu \sum_j \rho_j + \beta \nu \sum_j r_j} \big) = \mu \,.
\end{split}
\end{equation}
For the energy current there are more terms to be considered,
\begin{equation}
\begin{split}
\langle\mathcal{J}_{e,j}\rangle &= \big\langle V'(\rho_j)\sqrt{\rho_{j-1}\rho_j}\,U'(r_{j-1}) + \tfrac{1}{2} \sqrt{\rho_{j-1}\rho_{j+1}}\,\big(U(r_{j-1})U'(r_j) + U'(r_{j-1}) U(r_j) \big) \big\rangle \\[1ex]
&= -\beta^{-1} Z_{\mathrm{lt}}^{-1} \int \Big[ V'(\rho_j)\big(\partial_{r_{j-1}} \mathrm{e}^{-\beta H_\mathrm{lt}}\big) + \tfrac{1}{2}\sqrt{\rho_{j-1}\rho_{j+1}} \\
&\quad\times \Big\{ U(r_{j-1}) \tfrac{1}{\sqrt{\rho_{j}\,\rho_{j+1}}} \big(\partial_{r_j} \mathrm{e}^{-\beta H_\mathrm{lt}}\big) + U(r_{j}) \tfrac{1}{\sqrt{\rho_{j-1}\,\rho_{j}}} \big( \partial_{r_{j-1}} \mathrm{e}^{-\beta H_\mathrm{lt}} \big) \Big\} \Big] \, \mathrm{e}^{\beta\mu \sum_j \rho_j + \beta \nu \sum_j r_j} \\
&= \nu \, Z_{\mathrm{lt}}^{-1} \int \Big[ V'(\rho_j) + \tfrac{1}{2}\sqrt{\rho_{j+1}/\rho_j}\,U(r_j) + \tfrac{1}{2}\sqrt{\rho_{j-1}/\rho_j}\,U(r_{j-1}) \Big] \mathrm{e}^{-\beta (H_\mathrm{lt} - \mu \sum_j \rho_j - \nu \sum_j r_j)} \\
&= \mu\,\nu\,.
\end{split}
\end{equation}
Taking the limit to an infinitely high potential step, we conclude that the $H_\mathrm{lt}$-currents satisfy \eqref{eq:current_avr}. The zero $H$-currents would be obtained through a different limit, in which the potential barriers are removed.

%\bibliographystyle{unsrtmod}
%\bibliography{references}

\end{document}